\documentclass[aps,pra,nofootinbib,notitlepage,%
superscriptaddress,twocolumn]{revtex4-1}

\usepackage{amsmath}
\usepackage{amssymb}
\usepackage{natbib}
\usepackage{graphicx}
\usepackage{amssymb}
\usepackage{amsmath}
\usepackage{braket}
\usepackage{xcolor}
\usepackage{bm}
\usepackage{bbm}
\usepackage{hyperref}
\usepackage{siunitx}
\usepackage{dcolumn}

\newcolumntype{.}{D{x}{}{9}}

\newcommand{\calO}{\mathcal{O}}
\newcommand{\calH}{\mathcal{H}}

\newcommand{\dd}{\mathrm{d}}
\newcommand{\ii}{\mathrm{i}}
\newcommand{\ee}{\mathrm{e}}

\newcommand{\vecpt}{\vec p^{\,2}}

\usepackage{xcolor}
\usepackage{bbm}

\newcommand{\vp}{\mathrm{vp}}

\newcommand{\muH}{$\mu$H{}}
\newcommand{\muD}{$\mu$D{}}
\newcommand{\muHeThree}{$\mu\,{}^3$He}
\newcommand{\muHeFour}{$\mu\,{}^4$He}
\newcommand{\muCTwelve}{$\mu\,{}^{12}$C}
\newcommand{\muCThirteen}{$\mu\,{}^{13}$C}

\newcommand{\bcases}{\begin{cases}}
\newcommand{\ecases}{\end{cases}}
\newcommand{\bpm}{\begin{pmatrix}}
\newcommand{\epm}{\end{pmatrix}}
\newcommand{\bBm}{\begin{Bmatrix}}
\newcommand{\eBm}{\end{Bmatrix}}

\newcommand{\hsm}{\hspace{-2pt}}

\newcommand{\vsq}[1]{\vec #1 \hsm^2}
\newcommand{\crr}{\nonumber \\}

\newcommand{\ep}{\epsilon}

\newcommand{\half}{\frac{1}{2}}

\newcommand{\dbar}[2]{ \frac{\dd^#1 #2}{(2 \pi)^#1}}

\newcommand{\OC}{\mathrm{OC}}
\newcommand{\RC}{\mathrm{RC}}
\newcommand{\C}{\mathrm{C}}

\begin{document}

\title{Relativistic and Recoil Corrections to 
Vacuum Polarization in Muonic Systems:
Three--Photon Exchange, Gauge Invariance and Numerical Values}

\author{Gregory S. Adkins}
\affiliation{Department of Physics and Astronomy, Franklin \& Marshall College,
Lancaster, Pennsylvania 17604, USA}

\author{Ulrich D. Jentschura}
\affiliation{Department of Physics and LAMOR,
Missouri University of Science and Technology,
Rolla, Missouri 65409, USA}

\begin{abstract}
For an accurate theoretical description of muonic bound systems, it is crucial
to consistently treat relativistic and recoil corrections to vacuum
polarization. The one-loop vacuum-polarization effect is by far the dominant
quantum electrodynamic (QED) energy correction for bound muons, being of order
$\alpha (Z\alpha)^2 m_r$, where $\alpha$ is the fine-structure constant, $Z$ is
the nuclear charge number, and $m_r$ is the reduced mass.  Gauge invariance of
the relativistic and recoil corrections to vacuum polarization of order
$\alpha (Z\alpha)^4 m_r$ is investigated with respect to nonretarded and
standard, renormalized variants of Coulomb gauge. The invariance is shown 
after including three-photon exchange diagrams.  Our derivation is based on an adapted
form of Nonrelativistic Quantum Electrodynamics for bound muon systems
(NRQED$_\mu$), which is a version of NRQED where the hard scale is set at the
muon mass instead of the electron mass.  Updated values for the
gauge-independent corrections for one-muon ions with nuclear
charge numbers $Z = 1,2,6$ are presented.
\end{abstract}

\maketitle

\tableofcontents

\section{Introduction}
\label{sec1}

Vacuum-polarization effects are one of the predominant 
quantum electrodynamic (QED) energy corrections 
in simple bound systems.
The dominant (one-loop) vacuum-polarization 
correction to the photon propagator is 
due to the Uehling potential~\cite{Ue1935}.
The calculation of the leading vacuum polarization,
with modern calculational
techniques such as dimensional regularization,
has recently been recalled in Chap.~10 of 
Ref.~\cite{JeAd2022book} and in Ref.~\cite{LaJe2024}.
In two-body bound systems containing a muon bound to a nucleus,
electron vacuum polarization gives the leading contribution to the Lamb shift.
Recent work on closely related topics and some comprehensive discussions of 
energy corrections to two-body
muonic atoms and ions are contained in Refs. 
\cite{KaIvKa2013,KrEtAl2016,KaKoShIv2017,FrEtAl2017,ReDi2018,
DiEtAl2018,Re2018,DoEtAl2019,DoEtAl2020,AnHaPa2022,PaEtAl2024,LaJe2024}.

It has been known from the inception
of quantum electrodynamics that the 
choice of gauge
can crucially influence the degree of difficulty 
encountered in a bound-state calculation.
For example, the matching of the 
one-photon exchange scattering 
amplitude with the effective Breit Hamiltonian,
as described in Secs.~80--85 of 
Ref.~\cite{BeLiPi1982vol4},
is by far easiest in the Coulomb gauge,
where the time-time component of the 
photon propagator is free from 
retardation--that is, it mediates an instantaneous interaction.
In this gauge, the leading term is 
given exclusively by the Coulomb interaction,
that is, the time-time component of the 
photon propagator.
The spatial components of the 
Coulomb-gauge photon propagator, in the nonretardation 
approximation, give rise to the nonretarded Breit 
Hamiltonian (see also Chap.~11 of Ref.~\cite{JeAd2022book}).

If one insists on using a gauge other than Coulomb gauge
for QED bound-state calculations, then
heroic efforts are required even at tree 
level in order to show gauge invariance 
of the energy levels.
The literature is abundant with 
lengthy discussions of the 
matter~\cite{Lo1978,BaRe1978,FeFuHe1980i,FeFuHe1980ii,Li1990}.
One verifies the gauge invariance of energy levels
only after detailed and extensive considerations.

A subtle point arises in the consideration
of vacuum polarization.  We recall that the
time-time component of the Coulomb-gauge
bare photon propagator 
is completely free from retardation effects.  The
instantaneous nature of this binding Coulomb
interaction simplifies greatly the calculation
of relativistic corrections to energy levels, and
obviates the need for the consideration of retardation
corrections.  However, when vacuum polarization
effects are included to dress the
photon propagator, the time-time component of
the vacuum-polarization-corrected Coulomb propagator 
is not instantaneous.  We here refer to this gauge as the 
``renormalized Coulomb gauge'', although it is not
in fact a new gauge. Specifically, the ``renormalized Coulomb gauge'' 
is simply the Coulomb gauge when
vacuum polarization effects are included
and when expressed in terms of the physical charge.
In the instantaneous limit, this vacuum-polarization corrected
time-time photon leads at lowest order to the Uehling potential, which
corrects the energy levels by a term of order 
$\alpha (Z \alpha)^2 m_r$, where $\alpha$ is the fine-structure constant, 
$Z$ is the nuclear
charge number, and $m_r$ is the reduced mass. 
The dressed, non-instantaneous, propagator also gives rise to spurious
relativistic and recoil energy corrections of orders 
$\alpha (Z\alpha)^3 m_r$, $\alpha (Z\alpha)^4 m_r$ and higher orders.
However, a further
modification of the 
Coulomb gauge (see Refs.~\cite{Pa1996mu,Je2011pra,KaIvKo2012})
leads to an ``optimized Coulomb gauge'' 
where 
the time-time component 
of the photon propagator remains nonretarded,
and the
spurious contributions do not occur.

We here aim to verify the gauge invariance of 
relativistic and recoil energy corrections of orders 
$\alpha (Z\alpha)^3 m_r$ and $\alpha (Z\alpha)^4 m_r$,
between the ``renormalized'' and ``optimized'' Coulomb gauge
(see Sec.~\ref{sec4}). 
This invariance has been studied before 
(in Ref.~\cite{KaIvKo2012}) in the context of resolving
a difference between results for the $\alpha (Z \alpha)^4 m_r$ energy
shifts arising from one-loop vacuum polarization
found in Refs.~\cite{VePa2004,Bo2011preprint,Bo2012}
and \cite{Je2011pra}. We anticipate that the 
gauge invariance will only be obtained 
after considering two- and three-photon exchange,
with a retarded photon, 
corrected by vacuum-polarization,
crossing one or two Coulomb photons. In addition, we will also
generalize the
``optimized'' Coulomb gauge to all orders
in the vacuum-polarization insertion (see Sec.~\ref{sec3}).
Furthermore, after the demonstration of 
gauge independence of the energy corrections, 
we present (see Sec.~\ref{sec5}) 
values for bound states
having principal quantum number
$n=1-4$, 
of muonic hydrogen, muonic deuterium,
muonic helium (isotopes with mass numbers
3 and 4) and muonic carbon (isotopes with 
mass numbers 12 and 13). 
{\em A priori}, our calculations 
demonstrating gauge invariance of energy corrections
are carried out based on a formalism 
adapted to a spin-$\half$ bound particle
(muon) and a spin-$\half$ nucleus.
However, we shall realize that the only
terms required for gauge invariance
involve the spin-independent Coulomb exchange,
and the spin-independent exchange
of a magnetic photon.
The calculations are carried out within
a version of Nonrelativistic Quantum Electrodynamics
(NRQED) adapted to muonic bound systems,
which we refer to as NRQED$_\mu$ (see Sec.~\ref{sec2}).
Conclusions are reserved for Sec.~\ref{sec6}.

\section{Development of NRQED$_\mu$}
\label{sec2}

In order to calculate energy levels of 
muonic bound systems such as 
true muonium~\cite{JeSoIvKa1997,KaIvJeSo1998,KaJeIvSo1998,LaJi2018}
and muonic hydrogen~\cite{Pa1996mu,Je2011aop1,Je2011aop2}
one can use a version of NRQED (NRQED$_\mu$) where the hard cutoff
is set at the muon mass instead of the electron mass. 
We remember that, in ordinary NRQED (see Ref.~\cite{CaLe1986}
and Chap.~17 of Ref.~\cite{JeAd2022book}), 
the hard cutoff for NRQED is set at the 
electron mass scale, not the muon mass scale.
However, the presence of the heavier muon
in the bound system changes the mass hierarchy,
and the momentum scale of atomic binding (the ``soft'' scale) in muonic 
systems is close to the electron mass scale.
This means that, in NRQED$_\mu$, it is impossible
to expand the electron part of the Lagrangian 
about the nonrelativistic limit;
the electron remains a fully relativistic particle.

The Lagrangian of NRQED$_\mu$ has the form \cite{CaLe1986,HiLePaSo2013}
\begin{multline} 
\label{Lagrangian_outline}
\mathcal{L} = \sum_i \psi_i^\dagger 
\Bigl \{ \ii D_t + \frac{\vec D\, ^2}{2m_i} + \frac{\vec D\,^4}{8 m_i^3} 
+ c^{(i)}_F \frac{q_i}{2m_i} \vec \sigma_i \cdot \vec B
\\
+ c^{(i)}_D \frac{q_i}{8 m_i^2} 
\left ( \vec D \cdot \vec E - \vec E \cdot \vec D \right ) 
\\
+ c^{(i)}_S \frac{\ii q_i}{8 m_i^2} \vec \sigma_i \cdot 
\left ( \vec D \times \vec E - \vec E \times \vec D \right ) + \cdots \Bigr \} \psi_i \cr
+ \bar \Psi_e \left ( \ii \gamma^\mu D_\mu - m_e \right ) \Psi_e \\[3pt]
+ \mathrm{four\!-\!fermion \; contact \; terms} \cr
 + \mathrm{photon \; terms} + \mathrm{counter\; terms} \,,
\end{multline} 
where $\vec D$
is the covariant derivative, 
$\vec E$ and $\vec B$ are the quantized electric 
and magnetic fields, 
and the sum over the spin-$1/2$ fields $\psi_i$ extends 
over the participating fermions, typically a muon 
(charge $q_1=-\vert e \vert$, mass $m_1 = m_\mu$) and a 
positive particle such as a proton, a positive muon, or
a heavier nucleus ($q_2= Z \vert e \vert$),
so that $q_1 \, q_2 = -4 \pi Z \alpha$.    
The Pauli matrices are denoted as 
$\vec \sigma_i$, and the matching 
coefficients are $c^{(i)}_F $, $c^{(i)}_D $ and $c^{(i)}_S$.
The electron-positron  field $\Psi_e$ 
is fully relativistic, and has the usual QED electron Lagrangian factor.
The electron mass is denoted as $m_e$.
In the particle-antiparticle situation
(e.g., positronium and true muonium), there is the
possibility of virtual (and real) annihilation.  Both effects are represented
by contributions to the contact terms.  Also included in the contact terms
are the effects of hard scattering between the two participating
fermions.  The photon terms include the Maxwell term
$-\frac{1}{4} F_{\mu \nu} F^{\mu \nu}$ and terms representing the effect of
muonic and hadronic vacuum polarization. 
Here $F_{\mu\nu} = \partial_\mu A_\nu - 
\partial_\nu A_\mu$ is the field-strength tensor,
where $A_\mu$ is the four-vector potential.
Greek indices such as $\mu \in (0,1,2,3)$ and 
$\nu \in (0,1,2,3)$ are Lorentz indices.

The counter terms include
$-\frac{1}{4} (Z_3-1) F_{\mu \nu} F^{\mu \nu}$ that forces the electron vacuum
polarization function to vanish at zero momentum, as well as electron
wave-function and self-mass (self-energy) 
terms.  The charges and masses contained
in the NRQED$_\mu$ Lagrangian are the physical charges and masses.
The covariant derivatives 
are given in terms of the vector potential
as $D_\mu = \partial_\mu + \ii q A_\mu$, or 
\begin{subequations}
\label{def_cov_derivative} 
\begin{eqnarray}
D_t &=& \frac{\partial}{\partial t} + \ii q A^0, \\
\vec D &=& \vec \nabla - \ii q \vec A, 
\end{eqnarray}
\end{subequations}
where we recall that $(A^0, \vec A)$ is the four-vector potential,
$q$ is the charge of the particle,
and the electric and magnetic fields are defined in the usual way:
\begin{subequations}
\begin{eqnarray}
\vec E &=& -\vec \nabla A^0 - \frac{\partial \vec A}{\partial t}, \\
\vec B &=& \vec \nabla \times \vec A \,.
\end{eqnarray}
\end{subequations}

The nonrelativistic fermion quantum fields $\psi_i$ are 
two-component Pauli spinors.
Due to the nonrelativistic (NR) nature of the theory,  there is no
built-in unification of particles with their corresponding
antiparticles.  Furthermore, we reemphasize that
$\Psi_e$ is the (relativistic) electron-positron
field operator. The coefficients $c_F$, $c_D$,
$c_S$ are chosen so that NRQED$_\mu$ and the full QED agree when used to
calculate low-energy processes for which NRQED$_\mu$ is valid.  The low-energy
processes used in this ``matching'' procedure are typically the simplest
possible such as NR scattering processes, perhaps from a specified external
field. Details can be found in Chap.~17 of Ref.~\cite{JeAd2022book}.

The NRQED$_\mu$ Lagrangian shown in Eq.~\eqref{Lagrangian_outline} contains
just the first few terms of the complete Lagrangian, which contains an infinite
number of terms.  The remaining terms, 
which have origins in weak and strong interaction physics 
as well as high-energy QED,
all contain higher orders
in the inverses of heavy masses 
and give correspondingly smaller contributions to energies.  The
NRQED$_\mu$ Lagrangian is symmetric under translations, rotations, parity,
charge parity, and gauge transformations.  It is also invariant under Lorentz
transformations, but their representation is nonlinear (and non-obvious).  

For the purposes of the current investigation, the 
sum over $i$ in Eq.~\eqref{Lagrangian_outline}
extends over $i=1,2$, where $i=1$ is the muon,
and $i=2$ denotes the nucleus of charge number $Z$. In general,
the NRQED$_\mu$ theory has an upper cut-off at $m_\mu$ 
($m_\mu \sim 106 \, {\rm MeV}$); i.e., the contribution of
high-energy processes $E > m_\mu$ are represented by the matching coefficients.
Energy scales included in NRQED$_\mu$ include the soft 
($\alpha m_\mu \sim 0.77\,{\rm MeV}$) and ultrasoft 
($\alpha^2 m_\mu \sim 5.6\,{\rm keV}$) scales
and also the electron rest energy ($m_e \approx 0.511\,{\rm MeV}$).  We
regulate the divergences in NRQED$_\mu$ using dimensional regularization
($d=4-2\epsilon$), and use Coulomb gauge to define the photon propagator.  
(We note that the cross product and three-vector $\vec B$ are not defined
in a general number of dimensions.  In $d$ dimensions we replace 
$\epsilon^{a b c} \sigma_i^c$ by $\sigma_i^{a b} \equiv -\frac{\ii}{2} 
\left [ \sigma_i^a , \sigma_i^b \right ]$ and $\vec \sigma \cdot \vec B$ by
$-\frac{1}{2} \sigma_i^{a b} F_{a b}$.)
The consequence of having a relativistic electron is that
the photon propagator is modified by electron vacuum polarization.

\section{The Optimized Coulomb Gauge}
\label{sec3}

The electron is relativistic in NRQED$_\mu$, so electron vacuum polarization
corrects the photon propagator.  The vacuum-polarization tensor
\begin{equation}
\Pi_{\mu \nu}(k) = \Pi(k^2) \left ( k^2 g_{\mu \nu} - k_\mu k_\nu \right )
\end{equation}
is known to be independent of gauge (see Sec.~10.5 of \cite{We1995i}),
the same in Coulomb gauge as in the covariant gauges. 
We use the metric tensor 
$g_{\mu\nu}$ in ``West--Coast'' conventions where
$g_{0 0}=+1$ and $g_{i j}=-\delta_{i j}$ 
(Kronecker $\delta$)
and use $k^\mu$ to represent the photon 
momentum four-vector entering the 
vacuum-polarization loop. It follows that
$k^2 = (k^0)^2  - \vec k^2$ is negative for 
space-like momentum transfer (Coulomb photon). 
The renormalized
Coulomb gauge (RC) photon propagator is
\begin{eqnarray}
D^{\RC}_{\mu \nu}(k) &=& D^\C_{\mu \nu}(k) - D^\C_{\mu \alpha}(k) 
\Pi^{\alpha \beta}(k) D^\C_{\beta \nu}(k) \nonumber \\[3pt]
&\hbox{}& \hspace{-0.9cm} + D^\C_{\mu \alpha}(k) 
\Pi^{\alpha \beta}(k) D^\C_{\beta \gamma}(k) 
\Pi^{\gamma \kappa}(k) D^\C_{\kappa \nu}(k) - \cdots \, ,
\end{eqnarray}
where for the bare Coulomb (C) gauge photon propagator we have
\begin{equation} \label{bare_Coulomb}
D^\C_{\mu \nu}(k) =  \begin{pmatrix} 
\frac{1}{\vec k\,^2} & 0 \\ 0 & \frac{1}{k^2} 
\left ( \delta_{i j} - \hat k_i \hat k_j \right ) \end{pmatrix} \, .
\end{equation}
The sum for $D^{\RC}_{\mu \nu}(k)$ 
can be evaluated to all orders in the polarization
function and takes
the form of a geometric series:
\begin{equation}
\label{RC_photon_propagator}
D^{\RC}_{\mu \nu}(k) = \frac{1}{1-\Pi(k^2)} 
{ D}^\C_{\mu \nu}(k) \, .
\end{equation}
The value of the photon wave function counter-term factor $Z_3-1$ 
is defined so that $\Pi(0) = 0$. 
It is well known (see, e.g., Chap.~111 of 
Ref.~\cite{BeLiPi1982vol4},
and Ref.~\cite{Sc1970vol3})
that, based on the Cutkosky rules
(Ref.~\cite{Cu1960} and Sect.~7.3 of Ref.~\cite{PeSc1995}),
the vacuum-polarization insertion $\Pi(k^2)$ 
can be expressed as a dispersive integral.
Namely, its imaginary part is described
by the amplitude for the conversion of a virtual 
photon into (real) photons and fermions.
For one- and two-loop
diagrams with a fermion pair in the internal lines,
one integrates from the threshold $s = s_{\rm th} = (2 m_e)^2$
of pair production up to $s = \infty$.
The integration variable is conveniently 
parametrized as 
\begin{equation}
\lambda = \frac{2 m_e}{\sqrt{1-v^2}} \,,
\quad v \in (0,1) \,,
\quad 2 m_e < \lambda < \infty \,,
\end{equation}
and the dispersive integral has a representation 
\begin{equation}
\label{disp}
\Pi(k^2) = \int_0^1 \dd v \, F(v) \frac{k^2}{k^2-\lambda^2 + \ii \epsilon} \,.
\end{equation}
Up to two loops, the density function $F(v)$
has the perturbative (loop) expansion
\begin{equation}
F(v) = \frac{\alpha}{\pi} f_1(v) + 
\left ( \frac{\alpha}{\pi} \right )^2 f_2(v) + \calO(\alpha^3) \,.
\end{equation}
The one-loop function is
\begin{equation}
f_1(v) = \frac{v^2(1-\frac{v^2}{3})}{1-v^2} \, .
\end{equation}
The two-loop function $f_2(v)$ has been calculated 
and verified 
in Refs.~\cite{KaSa1955,Sc1970vol3,BaRe1973,EiSh2015,LaJe2024}.
For multi-loop (three-loop and beyond) effects, 
the dispersion relation~\eqref{disp} needs to be 
generalized according to 
Eq.~(111.8) of Ref.~\cite{BeLiPi1982vol4}.
At three loops and beyond, one may encounter a reduced threshold 
$s_{\rm th} = 0$ as opposed to $s_{\rm th} = 4 m_e^2$.
However, Eq.~\eqref{disp} remains valid 
if one replaces the expression $\frac{F(v)}{k^2 - \lambda^2 + \ii \epsilon}$ 
with $[- \frac{\dd s}{\dd v} \, \frac{{\rm Im} \, \Pi(s)}{\pi s}
\frac{1}{k^2 - s + \ii \epsilon}]$,
where $s = s(v)$ is a suitable parametrization
of the interval $s_{\rm th} < s < \infty$
[see also the comprehensive discussion
in Chap.~104 of Ref.~\cite{BeLiPi1982vol4} and 
Eq.~(104.11) of Ref.~\cite{BeLiPi1982vol4}].

The renormalized Coulomb gauge with the 
propagator (\ref{RC_photon_propagator}) is 
awkward for the calculation of muonic hydrogen
energy corrections because its time-time part
\begin{equation}
\label{RC_Coulomb_freq}
D^{\RC}_{00}(k) = \frac{1}{\vec k\,^2} +
\frac{1}{\vec k\,^2} \int_0^1 \dd v \, F(v) 
\frac{k^2}{k^2-\lambda^2} + \cdots \, ,
\end{equation}
involves the photon energy $k^0$, so in coordinate space it 
is not instantaneous.  This leads to anomalously large retardation contributions 
(namely, of order $\frac{\alpha}{\pi} (Z \alpha)^3$)  in the
one-photon exchange contribution
(in units of the reduced mass $m_r$ of the 
two-body bound system of muon and nucleus).
Furthermore, from two-photon exchange, as we shall
verify below, we also obtain an order 
$\frac{\alpha}{\pi} (Z \alpha)^3$ contribution.
An important part of the current investigation is to clarify the cancellation 
mechanism for the anomalous
contributions of orders $\frac{\alpha}{\pi} (Z \alpha)^3$ and
$\frac{\alpha}{\pi} (Z \alpha)^4$. 
In the instantaneous limit, 
the time-time component of $D^{\RC}_{\mu \nu}(k)$ becomes 
\begin{equation} \label{RC_Coulomb_prop}
D^{\rm R C}_{0 0}(k) \rightarrow \frac{1}{\vec k\,^2} + 
\int_0^1 \dd v \, F(v) \frac{1}{\vec k\,^2+\lambda^2} \, .
\end{equation}
 
A more advantageous gauge, which would avoid the anomalous 
contributions, would have $D^{\rm RC}_{00}(k) \vert_{k^0 \rightarrow 0}$ as its
 time-time component of the propagator. 
Indeed, the ideal gauge to use for NRQED$_\mu$ calculations has
this as its
time-time component and avoids non-zero time-space components.  Such a
gauge can be found, starting from the renormalized Coulomb gauge, by a gauge
change to an optimized Coulomb gauge (OC) with
the propagator given by
\begin{equation} 
\label{gauge_change}
D^{\rm O C}_{\mu \nu}(k) = D^{\RC}_{\mu \nu}(k) + b_\mu(k) k_\nu + k_\mu b_\nu(k) \, .
\end{equation}
The possibility to add such terms 
has been outlined in 
many places, including Sec.~76 of Ref.~\cite{BeLiPi1982vol4} 
and more recently in 
Sec.~9.5.2 of Ref.~\cite{JeAd2022book}.
To first order in $\Pi(k^2)$, $b_0$ can be found from
the requirement
\begin{multline} \label{OC_Coulomb_prop}
D^{\OC}_{0 0}(k) = \left ( 1 +  \Pi(k^2) \right ) 
D^{\C}_{0 0}(k) + 2 k_0 b_0(k) \\
=  \; D^\C_{0 0}(k) 
+ \int_0^1 \dd v \, F(v) \frac{1}{\vec k\,^2 + \lambda^2}  \,,
\end{multline}
where we have used the relation
$1/[1 -  \Pi(k^2)] \approx 1 +  \Pi(k^2)$
and thus
$D^{\RC}_{0 0}(k) \approx  \left [ 1 +  \Pi(k^2) \right ] 
D^{\C}_{0 0}(k)$.
The first-order expression for $b_0(k)$ is
\begin{equation} \label{eq_b0}
b_0(k) = - \frac{k_0}{2 \vec k\,^2} \int_0^1 \dd v \, F(v) 
\frac{\lambda^2}{\big ( \vec k\,^2 + \lambda^2 \big )
\left ( k^2-\lambda^2 \right )} \, .
\end{equation}
In order for the space-time off-diagonal terms $D^{\rm O C}_{i 0}(k)$ to vanish, it
is necessary that $b_i(k) = - k_i b_0(k)/k_0$, or
\begin{equation} \label{eq_bi}
b_i(k) = \frac{k_i}{2 \vec k\,^2} 
\int_0^1 \dd v \, F(v) \frac{\lambda^2}{\left ( k^2-\lambda^2 \right ) 
\big ( \vec k\,^2 + \lambda^2 \big )} \, .
\end{equation}
The first-order space-space components of the optimized Coulomb propagator are
then determined to be
\begin{multline}
D^{\rm O C}_{i j}(k) = D^\C_{i j}(k) \\ + 
\int_0^1 \dd v \, F(v) \frac{1}{k^2-\lambda^2} 
\left ( \delta_{i j} - \frac{k_i k_j}{\vec k\,^2 + \lambda^2 }  \right ) \, .
\end{multline}
This first-order 
optimized Coulomb gauge has been used in various calculations of energy
shifts for muonic atoms, 
including in Refs.~\cite{Pa1996mu,VePa2004,Je2011pra,KaIvKo2012}, 
and was called the ``C1eVP'' gauge in Ref.~\cite{KaIvKo2012}. 
Our Eqs.~\eqref{eq_b0} and~\eqref{eq_bi} are analogous to the expressions
given in Eq.~(15) of Ref.~\cite{KaIvKo2012},
but differ from those because of the different starting gauge.
The renormalized Coulomb gauge was
referred to as the ``C2eVP'' gauge in Ref.~\cite{KaIvKo2012}. 

In order to perform higher order calculations, it will be necessary to consider
the optimized Coulomb gauge to higher orders.  We insist that the optimized
Coulomb gauge has an instantaneous time-time propagator, so that
\begin{equation}
D^{\rm O C}_{0 0}(k) = \frac{1}{1-\Pi(-\vec k\,^2)} \frac{1}{\vec k\,^2} \, .
\end{equation}
The all-orders optimized Coulomb gauge comes from the renormalized Coulomb gauge according
to (\ref{gauge_change}), 
with $b_0$ chosen as
\begin{equation}
b_0(k) = \frac{1}{2 k_0 \vec k\,^2}  \frac{\Pi(-\vec k\,^2) - 
\Pi(k^2)}{\big [1-\Pi(-\vec k\,^2) \big ] \big [ 1-\Pi(k^2) \big ]} \, .
\end{equation}
We continue to require that the time-space components $D^{\rm O  C}_{i 0}(k)$
vanish, so that
\begin{equation}
b_i(k) = -\frac{k_i}{2 k_0^2 \vec k\,^2} 
\frac{\Pi(-\vec k\,^2) - \Pi(k^2)}{\big [1-\Pi(-\vec k\,^2) \big ] 
\big [ 1-\Pi(k^2) \big ]} \, .
\end{equation}

The all-orders form of the optimized Coulomb gauge propagator is
\begin{subequations} \label{OC_propagator}
\label{OC_gauge}
\begin{eqnarray}
D^{\rm O C}_{0 0}(k) &=& \frac{1}{1-\Pi(-\vec k\,^2)} \frac{1}{\vec k\,^2} \,, 
\\[3pt]
D^{\rm O C}_{i 0}(k) &=& D^{\OC}_{0 i}(k) = 0 \, , 
\\[3pt] 
D^{\rm O C}_{i j}(k) &=& \frac{1}{1-\Pi(k^2)} \frac{1}{k^2} 
\left ( \delta_{i j} - \frac{k_i k_j}{\vec k\,^2} \right ) \nonumber
\\
&\hbox{}& \hspace{-0.8cm} - \frac{k_i k_j}{k_0^2 \vec k\,^2} \frac{\Pi(-\vec k\,^2) - 
\Pi(k^2)}{\big [ 1-\Pi(k^2) \big ] \big [ 1-\Pi(-\vec k\,^2) \big ] } \, .
\end{eqnarray} 
\end{subequations}
An expansion through first order in $\Pi$ gives
\begin{multline}
D^{\rm O C}_{\mu \nu}(k) = \begin{pmatrix} \frac{1}{\vec k\,^2} & 0 \\ 0 & 
\frac{1}{k^2} \left ( \delta_{i j} - \frac{k_i k_j}{\vec k\,^2} \right ) 
\end{pmatrix}
+ \int_0^1 \dd v \, F(v)  \\
\times \begin{pmatrix} \frac{1}{\vec k\,^2+\lambda^2} & 0 \cr 0 & 
\frac{1}{k^2-\lambda^2} 
\left ( \delta_{i j} - \frac{k_i k_j}{\vec k\,^2+\lambda^2} 
\right) \end{pmatrix}+ \cdots \,  ,
\end{multline}
as above, and now the 
expansion to higher orders in $\Pi$ is possible when required.

\section{Verification of Gauge Invariance}
\label{sec4}

Gauge transformations in the context of Feynman diagrams are implemented by a
change in the photon propagator according to Eq.~\eqref{gauge_change}.
Such propagator changes, if applied to all photon propagators, cannot change
the result for the energy of a bound state. We recall 
(see Chaps.~16 and 17 of Ref.~\cite{JeAd2022book}) that the energy of a
two-body bound state appears as the position of a pole in the full 
two-to-two particle
Green function $G$ when considered as a function of the energy of the
two-particle system,
\begin{equation}
G(p_1',p_2';p_1,p_2) \rightarrow \ii \frac{\Psi(p_1',p_2') 
\bar \Psi(p_1,p_2)}{P^0-\sqrt{\vec P^2 + M^2}} \,,
\end{equation}
with $P=p_1+p_2 = p_1'+p_2'$, as 
$P^2 = (P^0)^2 - \vec P^2$ approaches the mass squared of a
two-body bound state.  The state-dependent mass of the bound state is
denoted as $M$, which is the sum of 
rest energies plus the binding energy.
A general proof that the bound-state energy levels are 
independent of gauge (the same in Coulomb gauge as in the covariant gauges)
has been given by Feldman {\it et al.}~\cite{FeFuHe1980i}.

In this section we will illustrate the gauge independence of energy levels of
muonic hydrogen; specifically, we aim to show that energy
corrections through order $\alpha(Z \alpha)^4 m_r$ calculated in the optimized
Coulomb gauge, agree with those calculated in the renormalized Coulomb gauge,
for contributions that are first order in vacuum polarization.

In the Coulomb gauge,
within the formalism of NRQED$_\mu$,
the graphs involving first-order vacuum polarization corrections
that contribute to energy levels at
order $\alpha(Z \alpha)^4$ are relatively 
few.  They are given as follows:
{\em (i)} vacuum-polarization (VP) corrected Coulomb exchange,
{\em (ii)} VP-corrected Coulomb exchange 
with a Darwin vertex on one side or the other,
{\em (iii)} VP-corrected Coulomb exchange with 
a spin-orbit vertex on one side or the other (part of the ``spin-orbit'' correction),
{\em (iv)} VP-corrected transverse photon exchange with 
convection vertices on each side (the ``magnetic'' correction),
{\em (v)} VP-corrected transverse photon exchange with one 
convection vertex and one Fermi vertex 
(another part of the ``spin-orbit'' correction'), and
{\em (vi)} VP-corrected transverse photon exchange with 
Fermi vertices on each side (the ``spin-spin'' correction).
The Feynman rules that can be used for the NRQED$_\mu$ vertices 
are given in Sec.~17.2 of Ref.~\cite{JeAd2022book}. 
The contribution of
{\em (i)} in the RC gauge is different from its contribution in the OC gauge
due to retardation.  For contributions {\em (ii)} and {\em (iii)} at order
$\alpha (Z \alpha)^4$, the instantaneous approximation is 
appropriate, so these contributions are the same in the RC gauge as in the OC gauge.
For the magnetic contribution {\em (iv)} we can also use the
instantaneous approximation, but this contribution differs between the RC gauge
and OC gauge due to the different $k_i k_j$ parts of the 
photon propagators.  Contributions {\em (v)} and {\em (vi)} are the
same in the two gauges, since both contain Fermi vertices that
vanish when multiplied by $k_i k_j$.  In addition, the graphs 
with one or two ladder Coulomb photons crossed by a
VP-corrected Coulomb photon also contribute at order
$\alpha (Z \alpha)^4$ in the RC gauge but not the OC gauge, as will be 
shown.

All of these contributions involve the exchange of a single 
VP-corrected
photon.  The ones denoted as ``VP-corrected Coulomb'' have the propagator
$\int_0^1 \dd v \, f_1(v) \frac{1}{\vec k\,^2+\lambda^2}$, and the ones
denoted as ``VP corrected transverse'' have propagator $\int_0^1 \dd v \, f_1(v)
\frac{1}{k^2-\lambda^2} \left ( \delta_{i j} - \frac{k_i k_j}{\vec
k\,^2+\lambda^2} \right )$ in the OC gauge.  In counting powers of $Z \alpha$, we consider
$\lambda = \frac{2 m_e}{\sqrt{1-v^2}}$ to be of order $\alpha$, because 
\begin{equation}
\label{beta}
\beta \equiv \frac{m_e}{\alpha m_r} \approx 0.737\,383\,7
\end{equation}
is certainly of order one for muonic hydrogen. 
Here, $m_\mu$ and $m_p$ denote the muon and proton masses, respectively,
and
\begin{equation}
m_r = \frac{m_\mu m_N}{m_\mu + m_N}
\end{equation}
is the reduced mass of the two-body bound system.
Here, $m_N$ is the nuclear mass, which is 
equal to the proton mass $m_p$ for muonic hydrogen.

Since $m_\mu$ sets the ``hard'' energy scale for NRQED$_\mu$, the
electron mass $m_e \approx \alpha m_\mu$ is considered to be soft, of
order $\alpha$.  In the renormalized Coulomb gauge, the corresponding propagators
are  $\int_0^1 \dd v \, f_1(v) \frac{k^2}{k^2-\lambda^2} \frac{1}{\vec
k\,^2}$ and $\int_0^1 \dd v \, f_1(v) \frac{1}{k^2-\lambda^2} \left (
\delta_{i j} - \frac{k_i k_j}{\vec k\,^2} \right )$.  

Now, let us discuss the additional terms that occur in the
renormalized Coulomb gauge.  The crucial difference between the optimized and
renormalized Coulomb gauges is the fact that the time-time propagator in the
latter involves $k_0$, which means that in coordinate space this propagation is
not instantaneous.  While the lowest order contribution of simple VP-corrected
Coulomb exchange in the two gauges is at order $\alpha (Z \alpha)^2$, in the OC
gauge the order $\alpha (Z \alpha)^2$ correction is the full contribution of
this graph, but in the RC gauge there are retardation corrections of orders
$\alpha (Z \alpha)^3$, $\alpha (Z \alpha)^4$, and all higher orders.  The other
corrections listed all contribute at order $\alpha (Z \alpha)^4$, but the
contribution of the magnetic correction is different in the two gauges.  Two
additional graphs contribute in the RC gauge (see Fig.~1): one with a single
Coulomb photon crossed by a VP-corrected Coulomb photon, and one with two
Coulomb photons in ladder configuration crossed by a VP-corrected Coulomb
photon.  The two-photon crossed graph gives a contribution starting at order
$\alpha (Z \alpha)^3$, while the three-photon crossed graph gives a
contribution starting at order $\alpha (Z \alpha)^4$.  All of this will be made
explicit in the calculations shown below.

\begin{figure*}[t]
\begin{center}
\begin{minipage}{0.99\linewidth}
\begin{center}
\includegraphics[width=0.98\linewidth]{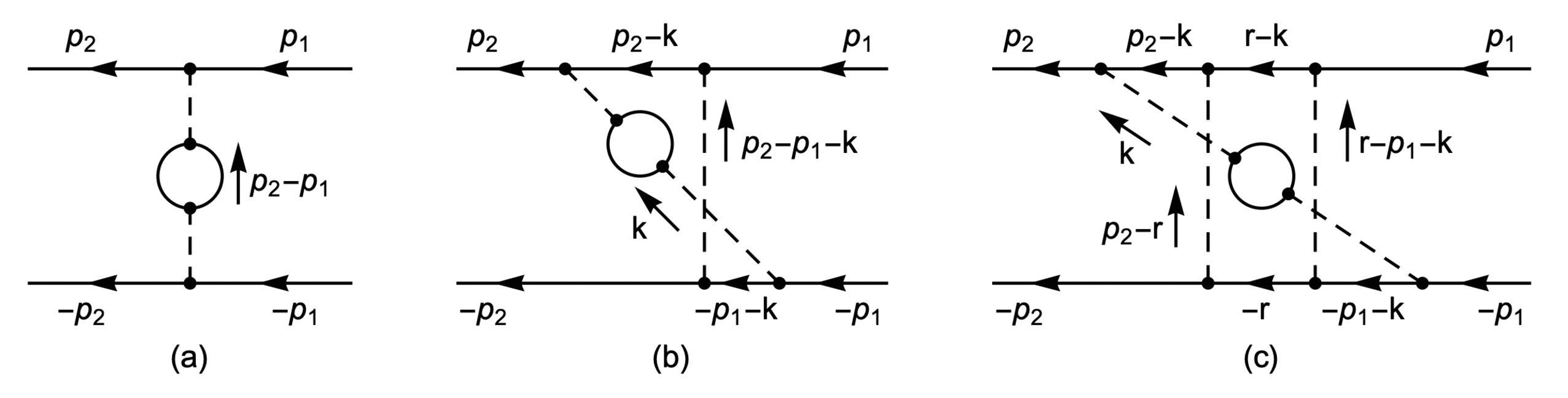}
\caption{\label{fig1} Graphs involving the exchange of a VP
corrected Coulomb photon crossing zero (a), one (b), or two (c) Coulomb photons
in the ladder configuration.  The top and bottom solid lines represent
the two constituent fermions.  The dashed lines represent Coulomb photons,
with propagator $D^\C_{00}$ as in Eq.~(\ref{bare_Coulomb}).
The VP corrected Coulomb lines have propagators given by
Eq.~(\ref{RC_Coulomb_freq}) in the renormalized Coulomb gauge, and by
Eq.~(\ref{OC_Coulomb_prop}) in the optimized Coulomb gauge.
In the renormalized Coulomb gauge,
the time-time component of the 
photon propagator acquires a residual photon energy 
dependence, which leads to retardation corrections.}
\end{center}
\end{minipage}
\end{center}
\end{figure*}

Our aim is to consider the contributions to energy levels at order $\alpha (Z
\alpha)^4$ coming from one-loop VP corrections in the renormalized Coulomb gauge
one by one and note 
the differences from the optimized Coulomb gauge.  We will
show that the differences add up to zero.
We start with the VP-corrected one-photon-exchange graph
[see Fig.~\ref{fig1}(a)].  
The energy shift is (see Refs.~\cite{Ad2018,JeAd2022book})
\begin{multline}
\label{single_Coulomb_exchange}
\Delta E_C = \ii 
\frac{\alpha}{\pi} 
\int \dbar{4}{p_2} \, \dbar{4}{p_1} \, 
\bar \Psi(p_2) (-\ii q_1) 
\int_0^1 \dd v \, f_1(v) 
\\
\times 
\frac{\ii}{\vec k\,^2} 
\frac{k^2}{k^2-\lambda^2} (-\ii q_2) \Psi(p_1) 
= (-4 \pi Z \alpha)  \frac{\alpha}{\pi}
\\
\times \int_0^1 \dd v \, f_1(v) 
\int \dbar{3}{p_2} \, \dbar{3}{p_1} 
\psi^\dagger(\vec p_2) \frac{R(\vec p_2,\vec p_1)}{\vec k\,^2} \psi(\vec p_1),
\end{multline}
where $R(\vec p_2,\vec p_1)$ is defined in Eq.~\eqref{defR}
and the wave functions are those of 
the NRQED$_\mu$ Bethe-Salpeter equation
(see Sec.~17.2.3 of Ref.~\cite{JeAd2022book}): \break
\begin{eqnarray}
\Psi(p) &=& -\ii S_1(p) S_2(-p) s(\vec p\,)^{-1} \psi(\vec p\,) \,, 
\crr
\bar \Psi(p) &=& -\ii \psi^\dagger(\vec p\,) s(\vec p\,)^{-1} S_1(p) S_2(-p)  \, ,
\end{eqnarray}
with
\begin{subequations}
\begin{align}
s(\vec p) =& \; \frac{1}{E-\frac{\vec p\,^2}{2 m_r} + \ii \ep} \,, 
\\
S_i(p) =& \;  \frac{1}{\xi_i E+p_0-\frac{\vec p\,^2}{2 m_i} + \ii \ep} \, .
\end{align}
\end{subequations}
for a bound state of energy $E$, where $\xi_i = m_i/M$ with $M = (m_1+m_2)$.
We have suppressed the bound-state quantum numbers, 
while $\psi(\vec p\,)$ is the
usual Schr\"odinger--Pauli (two-component) wave function. 
The energy integrals are contained
in the quantity $R(\vec p_2,\vec p_1)$:
\begin{widetext}
\begin{eqnarray}
\label{defR}
R(\vec p_2,\vec p_1) &=& (-\ii)^2 s^{-1}(\vec p_2) s^{-1}(\vec p_1) 
\int \frac{d p_{2 0}}{2\pi} \, \frac{dp_{1 0}}{2\pi} \, k^2  S_1(p_2) S_2(-p_2)
\big [ (k_0-\Omega+\ii \ep) (k_0+\Omega-\ii \ep) \big ]^{-1}
S_1(p_1) S_2(-p_1) \crr
&=& \frac{2(\Omega-E)(\vec k\,^2-\Omega E)+(\vec k\,^2+\Omega^2-2 \Omega E) 
\left ( \frac{\vec p_1^{\,2}}{2 m_r} + \frac{\vec p_2^{\,2}}{2 m_r} \right ) + 
2 \Omega \left ( \frac{\vec p_1^{\,2}}{2 m_1} + 
\frac{\vec p_2^{\,2}}{2 m_2} \right) 
\left ( \frac{\vec p_1^{\,2}}{2 m_2}+\frac{\vec p_2^{\,2}}{2 m_1} \right )}
{2 \Omega \left ( \Omega - E + 
\frac{\vec p_1^{\,2}}{2 m_2} + \frac{\vec p_2^{\,2}}{2 m_1} \right ) 
\left ( \Omega - E + \frac{\vec p_1^{\,2}}{2 m_1} + 
\frac{\vec p_2^{\,2}}{2 m_2}  \right ) } \, ,
\end{eqnarray}
\end{widetext}
where $\vec k \equiv \vec p_2-\vec p_1$ and 
we define
\begin{equation}
\Omega =  \sqrt{\vec k\,^2+\lambda^2 } \, .
\end{equation}
 (We will use $\vec k$ for the 
three-momentum of the VP-corrected
photon uniformly through all our calculations.) 
 We perform the energy
integrals using the residue theorem by closing contours in either the upper or
lower half plane.  We let all infinitesimals $\ep$ have distinct positive
values.  Choices must be made for which half-plane (upper or lower) to close 
the integration contour 
for each energy integral and for the ordering of the $\ep$'s.  All choices lead
to the same result.

Upon expansion, assuming that $\vec p_1$, $\vec p_2$, $\vec k$, and $\Omega$
are soft (of order $Z \alpha m_r$), and the bound state energy $E$ is ultrasoft 
(of order $(Z \alpha)^2 m_r$), we obtain
\begin{multline} 
\label{expan_of_R_for_C}
R(\vec p_2,\vec p_2) = \frac{\vec k\,^2}{\Omega^2} - 
\frac{\lambda^2}{2 \Omega^3} \left ( s^{-1}(\vec p_1) + s^{-1}(\vec p_2) \right ) 
\\
- \frac{\lambda^2}{2 \Omega^4} 
\left \{ 2 s^{-1}(\vec p_2) s^{-1}(\vec p_1) + 
\frac{(\vsq{p_2}-\vsq{p_1})^2}{4} 
\left ( \frac{1}{m_1^2} + \frac{1}{m_2^2} \right ) \right \} 
\\ + \cdots \, .
\end{multline}
The first term in the expansion gives the leading 
$\calO[\alpha (Z \alpha)^2 m_r]$ energy correction,
\begin{multline}
\Delta E_{C0} = \frac{\alpha}{\pi} \int\limits_0^1 \dd v f_1(v) 
\\
\times \int \dbar{3}{p_2} \dbar{3}{p_1} \psi^\dagger(\vec p_2) 
\frac{-4 \pi Z \alpha}{\Omega^2} \psi(\vec p_1)
\\
= \frac{\alpha}{\pi} \int_0^1 \dd v \, f_1(v) \int d^3 x \, 
\psi^\dagger(\vec x) \frac{-Z \alpha e^{-\lambda r}}{r} \psi(\vec x) \, .
\end{multline}
The second and third terms in Eq.~\eqref{expan_of_R_for_C}
are of orders $\alpha (Z \alpha)^3 m_r$ and 
$\alpha (Z \alpha)^4 m_r$, respectively.  While the first term is present in both
the OC and RC gauges, the second and third terms are retardation corrections
that only appear in the RC gauge.

The next term where a difference between OC and RC results 
appears at $\calO[\alpha (Z \alpha)^4 m_r]$ is 
the magnetic correction, which comes from transverse
photon exchange with convection vertices on both the muon and proton lines.
The difference comes from the $k_i k_j$ terms of the transverse photon
propagator, which is different in the two gauges.  This difference is
\begin{multline}
\delta D_{i j}(k) = D^{\RC}_{i j}(k) - D^{\OC}_{i j}(k) 
\\
= 
\frac{\alpha}{\pi} \int_0^1 \dd v \, \frac{f_1(v)}{k^2-\lambda^2} 
\left[ \left ( \delta_{i j} - \frac{k_i k_j}{\vec k\,^2} \right ) - 
\left ( \delta_{i j} - \frac{k_i k_j}{\vec k\,^2+\lambda^2} 
\right ) \right] \\
= \frac{\alpha}{\pi} 
\int_0^1 \dd v \, f_1(v) \frac{1}{k^2-\lambda^2} 
\frac{-\lambda^2 k_i k_j}{\vec k\,^2 (\vec k\,^2+\lambda^2) } 
\\
\rightarrow  \frac{\alpha}{\pi} \int_0^1 \dd v \, f_1(v) 
\frac{\lambda^2 k_i k_j}{\vec k\,^2 (\vec k\,^2+\lambda^2)^2 } \,,
\end{multline}
to lowest order, which is the only order that 
contributes at $\calO[\alpha (Z \alpha)^4 \, m_r]$. 
\begin{widetext}
The corresponding change in the magnetic energy correction is
\begin{eqnarray} \label{result_M}
\delta \Delta E_M &=& \ii \int \dbar{4}{p_2} \, 
\dbar{4}{p_1} \, \bar \Psi(p_2) 
\frac{\ii q_1}{2 m_1} (p_2+p_1)_i \, \ii \,
\delta D_{i j}(k) \frac{\ii q_2}{2 m_2} (-p_2-p_1)_j \Psi(p_1) \crr
&=&(-4 \pi Z \alpha) \frac{\alpha}{\pi} 
\int_0^1 \dd v \, f_1(v) \int \dbar{3}{p_2} \, \dbar{3}{p_1} \psi^\dagger(\vec p_2) 
\frac{\lambda^2 }{\vec k\,^2 
(\vec k\,^2+\lambda^2)^2 } 
\frac{(\vsq{p_2}-\vsq{p_1})^2}{8} 
\left ( \frac{-2}{m_1 m_2} \right ) \psi(\vec p_1) \, ,
\end{eqnarray}
which combines nicely with the last term in 
Eq.~\eqref{expan_of_R_for_C}; 
eventually, it cancels against a term from the crossed two-photon graph.

The crossed two-photon graph of interest is the one with a regular Coulomb
photon crossed by a VP-corrected Coulomb photon, 
shown in Fig.~\ref{fig1}(b). The corresponding energy
correction in the RC gauge is
\begin{eqnarray}
\Delta E_{CC} &=& 2 \, \ii \,\frac{\alpha}{\pi} 
\int_0^1 \dd v \, f_1(v) \int \dbar{4}{p_2} \, 
\dbar{4}{p_1} \, \dbar{4}{k} \, \bar \Psi(p_2) (-\ii q_1) \, 
\ii \, S_1(p_2-k) (-\ii q_1) \crr
&\hbox{}& \hspace{0.2cm} \times 
\frac{\ii}{(\vec k - \vec q \,)^2} \frac{\ii}{\vec k\,^2} 
\frac{k^2}{k^2-\lambda^2} \, (-\ii q_2) \, \ii \, 
S_2(-p_1-k) (-\ii q_2) \Psi(p_1) \crr
&=& -2 (-4 \pi Z \alpha)^2 \, \frac{\alpha}{\pi} 
\int_0^1 \dd v \, f_1(v) \int \dbar{3}{p_2} \, \dbar{3}{p_1} \, \dbar{3}{k} \, 
\psi^\dagger(\vec p_2) \,
\frac{R(\vec p_2,\vec k,\vec p_1)}{(\vec k-\vec q)^2 \vec k\,^2} 
\psi(\vec p_1) \, ,
\end{eqnarray}
where we have used $p_1 = | \vec p_1 |$, $p_2 = | \vec p_2 |$,
$\vec q = \vec p_2-\vec p_1$, and the factor of two comes because
there are two equal crossed graphs of this type. The energy integrals give
\begin{eqnarray}
R(\vec p_2,\vec k,\vec p_1) &=& (-\ii)^3 s^{-1}(\vec p_2) 
s^{-1}(\vec p_1) \int \frac{d p_{2 0}}{2 \pi} \, 
\frac{d p_{1 0}}{2 \pi} \, \frac{d k_0}{2 \pi} \, S_1(p_2) S_2(-p_2) S_1(p_2-k) 
\frac{k^2}{k^2-\lambda^2} S_2(-p_2-k) S_1(p_1) S_2(-p_1) \crr
&=& -\frac{\lambda^2}{2 \Omega 
\left( \Omega -E+\frac{\vec p_1^{\,2}}{2 m_1} + 
\frac{(\vec p_1+\vec k)^2}{2 m_2} \right ) 
\left( \Omega - E+\frac{\vec p_2^{\,2}}{2 m_2} + 
\frac{(\vec p_2-\vec k)^2}{2 m_1} \right ) } \, .
\end{eqnarray}
The soft expansion of this interaction
kernel $R$ gives 
\begin{eqnarray} \label{expan_R_CC}
R(\vec p_2,\vec k,\vec p_1) &=&- \frac{\lambda^2}{2 \Omega^3} - 
\frac{1}{2 \Omega^4} \left 
\{ 2 E -\frac{\vsq{p_1}}{2 m_1} -
\frac{(\vec p_1+\vec k)^2}{2 m_2} - 
\frac{(\vec p_2-\vec k)^2}{2 m_1} - 
\frac{\vsq{p_2}}{2 m_2} \right \} + \cdots \crr
&=&- \frac{\lambda^2}{2 \Omega^3} - 
\frac{\lambda^2}{2 \Omega^4} 
\left \{ s^{-1}(\vec p_1) + s^{-1}(\vec p_2) +
\frac{2 \vec p_2 \cdot \vec k-\vec k\,^2}{2 m_1} - 
\frac{2 \vec p_1 \cdot \vec k+\vec k\,^2}{2 m_2} \right \} + \cdots \, .
\end{eqnarray}
The leading term here gives an anomalously large $\calO[\alpha (Z \alpha)^3 m_r]$
contribution that cancels the $\calO[\alpha (Z \alpha)^3 m_r]$ contribution from the
retardation correction to single photon exchange, as we will show.  

Before exhibiting the cancellation
mechanism, we first set up the three-photon-exchange graph
with the momenta denoted as in Fig.~\ref{fig1}(c).
The three-photon-exchange graph with two Coulomb photons in ladder
configuration crossed by a VP-corrected Coulomb photon, in the RC gauge, 
leads to an energy shift of 
\begin{eqnarray}
\Delta E_{CCC} &=& 2 \ii \frac{\alpha}{\pi} 
\int_0^1 \dd v \, f_1(v) \int \dbar{4}{p_2} \, 
\dbar{4}{p_1} \, \dbar{4}{k} \, \dbar{4}{r} \, 
\bar \Psi(p_2) (-\ii q_1) \ii S_1(p_2-k) (-\ii q_1) \ii S_1(r-k) (-\ii q_1) \crr
&\hbox{}& \hspace{0.2cm} \times \frac{\ii}{(\vec p_2-\vec r)^2} 
\frac{\ii}{( \vec r - \vec p_1 - \vec k \, )^2 } 
\frac{\ii}{\vec k\,^2} \frac{k^2}{k^2-\lambda^2} 
(-\ii q_2) \ii S_2(-r) (-\ii q_2) \ii S_2(-p_1-k) (-\ii q_2) \Psi(p_1) \crr
&=& 2 (-4 \pi Z \alpha)^3 \frac{\alpha}{\pi} 
\int_0^1 \dd v \, f_1(v) \int \dbar{3}{ p_2} \, 
\dbar{3}{ p_1} \, \dbar{3}{ k} \, \dbar{3}{ r} \, 
\psi^\dagger(\vec p_2) \frac{R}{(\vec p_2-\vec r)^2 
(\vec r - \vec p_1 - \vec k \, )^2 \vec k\,^2} \psi(\vec p_1) \, .
\end{eqnarray}
The energy integrals give
\begin{eqnarray}
R &=& (-\ii)^4 s^{-1}(\vec p_2) s^{-1}(\vec p_1) \int \frac{d p_{2 0}}{2 \pi} \, 
\frac{d p_{1 0}}{2 \pi} \, \frac{d k_0}{2 \pi} \, \frac{d r_0}{2 \pi}  \, 
S_1(p_2) S_2(-p_2) S_1(p_2-k) S_1(r-k) \crr
&\hbox{}& \hspace{1.2cm} \times 
\frac{k^2}{k^2-\lambda^2} S_2(-r) S_2(-p_1-k) S_1(p_1) S_2(-p_1) \crr
&=& -\frac{\lambda^2}{2 \Omega \left ( \Omega -E+\frac{\vec{p_1}}{2 m_1} + 
\frac{(\vec p_1+\vec k)^2}{2 m_2} \right ) 
\left ( \Omega - E+\frac{\vsq{p_2}}{2 m_2} + 
\frac{(\vec p_2-\vec k)^2}{2 m_1} \right ) 
\left ( \Omega - E + \frac{\vec r^2}{2 m_2} + \frac{(\vec r-\vec k)^2}{2m_1} \right ) }
= - \frac{\lambda^2}{2 \Omega^4} + \cdots 
\end{eqnarray}
after soft expansion.  The corresponding energy contribution, which is
$\calO[\alpha (Z \alpha)^4 m_r]$, is given by the expression
\begin{equation} \label{DeltaE_CCC0}
\Delta E_{CCC0}
= - (-4 \pi Z \alpha)^3 \frac{\alpha}{\pi} 
\int_0^1 \dd v \, f_1(v) \int \dbar{3}{ p_2} \, \dbar{3}{ p_1} \, 
\dbar{3}{ k} \, \dbar{3}{ r} \, 
\psi^\dagger(\vec p_2) \frac{\lambda^2}{(\vec p_2-\vec r)^2 
(\vec r - \vec p_1 - \vec k \, )^2 \vec k\,^2 \Omega^4} \psi(\vec p_1) \, .
\end{equation}
Graphs with a transverse photon exchange with one or two Fermi vertices (which
are proportional to $\sigma_{i m} k_m$) vanish when a Fermi vertex is
multiplied by $k_i$, as in $\delta D_{i j}(k)$, so they contribute the same at
$\calO[\alpha (Z \alpha)^4 m_r]$ in the RC gauge as in the OC gauge.

We are now prepared to demonstrate the cancellations that result in the gauge
invariance of the vacuum-polarization 
energy correction through order $\alpha (Z \alpha)^4 m_r$. 
The leading term $\Delta E_{C0}$ from
the VP-corrected one-photon exchange is the 
same as in the OC gauge. Thus, in the RC gauge, we shall need to consider the 
next-to-leading term $\Delta E_{C1}$ and the 
next-to-next-to-leading term $\Delta E_{C2}$ 
from one-photon exchange,
as well as the leading term $\Delta E_{CC0}$ and next-to-leading term $\Delta E_{CC1}$
from two-photon exchange given in Fig.~\ref{fig1}(b),
while the three-photon term from Fig.~\ref{fig1}(c)
only needs to be considered in leading order,
\begin{equation}
\Delta E_C \approx \Delta E_{C0} + \Delta E_{C1} + \Delta E_{C2} \,,
\qquad
\Delta E_{CC} \approx \Delta E_{CC0} + \Delta E_{CC1} \,,
\qquad
\Delta E_{CCC} \approx \Delta E_{CCC0} \,.
\end{equation} 
We will make extensive use of the momentum space
Schr\"odinger equation
and its adjoint
\begin{subequations}
\begin{eqnarray}
\left ( E - \frac{\vec p\,^2}{2 m_r} \right ) \psi(\vec p \, ) &=&
s^{-1}(\vec p \, ) \psi(\vec p \, ) = \int \dbar{3}{ r }
\frac{-4 \pi Z \alpha}{(\vec p -\vec r \, )^2} \psi(\vec r \, ) \,, \\[3pt]
\psi^\dagger(\vec p \, ) \left ( E - \frac{\vec p\,^2}{2 m_r} \right ) &=& 
\psi^\dagger(\vec p \, ) s^{-1}(\vec p \, ) = 
\int \dbar{3}{ r} \psi^\dagger(\vec r \, ) 
\frac{-4 \pi Z \alpha}{(\vec r -\vec p \, )^2} \, .
\end{eqnarray}
\end{subequations}

There are two contributions at order $\alpha (Z \alpha)^3 m_r$, coming from the
first recoil correction to the single VP-corrected photon exchange and the
leading term from two-photon exchange.  These are
\begin{subequations}
\begin{eqnarray}
\Delta E_{C 1} &=& (-4 \pi Z \alpha) \frac{\alpha}{\pi} 
\int_0^1 \dd v \, f_1(v) \int \dbar{3}{ p_2} \, 
\dbar{3}{ p_1} \psi^\dagger(\vec p_2) \frac{1}{\vec k\,^2} 
\left \{ - \frac{\lambda^2}{2 \Omega^3} 
\left ( s^{-1}(\vec p_1) + s^{-1}(\vec p_2) \right ) \right \} 
\psi(\vec p_1) \, ,\\[3pt]
\Delta E_{CC 0} &=& -2 (-4 \pi Z \alpha)^2 \, \frac{\alpha}{\pi} \,
\int_0^1 \dd v \, f_1(v) \int \dbar{3}{ p_2} \, 
\dbar{3}{ p_1} \, \dbar{3}{ k} \, 
\psi^\dagger(\vec p_2) 
\frac{1}{(\vec p_2-\vec p_1-\vec k)^2 \vec k\,^2} 
\left \{ - \frac{\lambda^2}{2 \Omega^3} \right \} \psi(\vec p_1) \, ,
\end{eqnarray} 
\end{subequations}
respectively, where $\vec k = \vec p_2 - \vec p_1$ in $\Delta E_{C 1}$,
while $\vec k$ is an integration variable in 
$\Delta E_{CC 0}$. We apply the Schr\"odinger
equation to $\Delta E_{CC 0}$ by integrating over $\vec p_2$ and over 
$\vec p_1$ and averaging the results:

\begin{eqnarray} 
\label{cancel1}
\Delta E_{CC0} &=& - (-4 \pi Z \alpha) \frac{\alpha}{\pi} 
\int_0^1 \dd v \, f_1(v) 
\int \dbar{3}{ p_1} \, \dbar{3}{ k} \, 
\psi^\dagger(\vec p_1+\vec k) s^{-1}(\vec p_1+\vec k) 
\frac{1}{ \vec k\,^2} 
\left \{ - \frac{\lambda^2}{2 \Omega^3} \right \} \psi(\vec p_1) \crr
&\hbox{}& \hspace{0.05cm} - (-4 \pi Z \alpha) \frac{\alpha}{\pi} 
\int_0^1 \dd v \, f_1(v) \int \dbar{3}{ p_2}  \, 
\dbar{3}{ k} \, \psi^\dagger(\vec p_2) 
\frac{1}{ \vec k\,^2} 
\left \{ - \frac{\lambda^2}{2 \Omega^3} \right \} 
s^{-1}(\vec p_2 - \vec k) \psi(\vec p_2-\vec k) = -\Delta E_{C1} \,.
\end{eqnarray}
The latter equality can be shown 
upon changing the $\vec k$ integration variable in each integral according to
$\vec k \rightarrow \vec p_2 - \vec p_1$. The two terms in
(\ref{cancel1}) then simplify, leading 
to the result $\Delta E_{CC0} + \Delta E_{C 1} = 0$.  We
conclude that the order $\alpha (Z \alpha)^3 m_r$ terms from retardation in
one-photon-exchange and from the exchange of two crossed photons exactly
cancel.

We now show that the terms of order $\alpha (Z \alpha)^4 m_r$ 
also cancel, with contributions 
coming from one-, two-, and three-photon exchange and also
from the magnetic term.  The plan is to use the Schr\"odinger equation to
reduce the $3D$- and $4D$-dimensional integrals of $\Delta E_{CC1}$ and $\Delta
E_{CCC0}$ to $2D$-dimensional integrals (where $D=3$ here), then combine with $\delta \Delta E_M$
and $\Delta E_{C2}$ of the magnetic and single Coulomb terms.  Starting with
three-photon-exchange $\Delta E_{CCC0}$ of 
Eq.~\eqref{DeltaE_CCC0}, we integrate
over $\vec p_2$ and $\vec p_1$ to find
\begin{eqnarray} 
\label{result_CCC0}
\Delta E_{CCC0} &=& - (-4 \pi Z \alpha) 
\frac{\alpha}{\pi} 
\int_0^1 \dd v \, f_1(v) 
\int  \dbar{3}{ k} \, \dbar{3}{ r} \, 
\psi^\dagger(\vec r \, ) 
s^{-1}(\vec r \,) 
\frac{\lambda^2}{ \vec k\,^2 \Omega^4}s^{-1}(\vec r- \vec k \, )  \psi(\vec r - \vec k\,) \, \crr
&=& - (-4 \pi Z \alpha) 
\frac{\alpha}{\pi} 
\int_0^1 \dd v \, f_1(v) 
\int  \dbar{3}{ p_2} \, \dbar{3}{ p_1} \, 
\psi^\dagger(\vec p_2) 
s^{-1}(\vec p_2) s^{-1}(\vec p_1) 
\frac{\lambda^2}{ \vec k\,^2 \Omega^4} \psi(\vec p_1) \, ,
\end{eqnarray} 
where we have changed integration variables 
according to $\vec r \rightarrow \vec p_2$, $\vec k \rightarrow \vec p_2-\vec p_1$, 
and continue to use 
$\Omega^2 = \vec k\,^2 + \lambda^2$. 
Furthermore, we still use 
$\vec k \rightarrow \vec p_2-\vec p_1$ for convenience in the second line.
Similarly, the order $\alpha (Z \alpha)^4 m_r$ contribution
from the two-photon-exchange graph in Fig.~\ref{fig1}(b) is
\begin{eqnarray}
\Delta E_{CC1} &=&  (-4 \pi Z \alpha)^2 \frac{\alpha}{\pi} 
\int_0^1 \dd v \, f_1(v) \int  \dbar{3}{ p_2} \, \dbar{3}{ p_1} \, \dbar{3}{ k} \crr
& & \times \psi^\dagger(\vec p_2) 
\left \{ \left ( s^{-1}(\vec p_1) - 
\frac{2 \vec p_1 \cdot \vec k+\vec k\,^2}{2 m_2} \right ) 
+ \left ( s^{-1}(\vec p_2)+ 
\frac{2 \vec p_2 \cdot \vec k-\vec k\,^2}{2 m_1} \right ) \right \}  
\frac{\lambda^2}{ (\vec p_2-\vec p_1 - \vec k)^2 
\vec k\,^2 \Omega^4} \psi(\vec p_1) \crr
&=&  (-4 \pi Z \alpha) \frac{\alpha}{\pi} \int_0^1 \dd v \, f_1(v) 
\Bigg \{ \int  \dbar{3}{ k}  \, 
\dbar{3}{ p_1} \, \psi^\dagger(\vec p_1 + \vec k) s^{-1}(\vec p_1+\vec k) 
\left ( s^{-1}(\vec p_1) - 
\frac{2 \vec p_1 \cdot \vec k+\vec k\,^2}{2 m_2} \right ) 
\frac{\lambda^2}{ \vec k\,^2 \Omega^4} \psi(\vec p_1) \crr
& & + \int  \dbar{3}{ p_2}  \, 
\dbar{3}{ k} \, \psi^\dagger(\vec p_2) 
\left ( s^{-1}(\vec p_2)+ \frac{2 \vec p_2 \cdot \vec k-\vec k\,^2}{2 m_1} \right )
\frac{\lambda^2}{ \vec k\,^2 \Omega^4} 
s^{-1}(\vec p_2-\vec k) \psi(\vec p_2-\vec k) \Bigg \} \, .
\end{eqnarray}
We now use the variable changes
$(\vec k, \vec p_1) \to (\vec p_2-\vec p_1, \vec p_1)$ 
in the first term and 
$(\vec k, \vec p_2) \to (\vec p_2-\vec p_1, \vec p_2)$ 
in the second to write this as
\begin{multline} \label{CC1_form3}
\Delta E_{CC1} =  (-4 \pi Z \alpha) \frac{\alpha}{\pi} 
\int_0^1 \dd v \, f_1(v) \int  \dbar{3}{ p_2} \, \dbar{3}{ p_1}  \, 
\psi^\dagger(\vec p_2) \frac{\lambda^2}{ \vec k\,^2 \Omega^4} 
\psi(\vec p_1)
\biggl\{ 2 s^{-1}(\vec p_2) s^{-1}(\vec p_1) - 
\left ( E - \frac{\vsq{p_2}}{2 m_r} \right ) 
\frac{(\vsq{p_2}-\vsq{p_1})}{2 m_2} \\
+ \frac{(\vsq{p_2}-\vsq{p_1})}{2 m_1} 
\left ( E - \frac{\vsq{p_1}}{2 m_r} \right )  \biggr\} 
= (-4 \pi Z \alpha) \frac{\alpha}{\pi} 
\int_0^1 \dd v \, f_1(v) \int  \dbar{3}{ p_2} \, 
\dbar{3}{ p_1}  \, \psi^\dagger(\vec p_2) 
\frac{\lambda^2}{ \vec k\,^2 \Omega^4}  \psi(\vec p_1) 
\\
\times \left\{ 2 s^{-1}(\vec p_2) s^{-1}(\vec p_1) - 
E \, (\vsq{p_2}-\vsq{p_1}) 
\left( \frac{1}{2 m_2}-\frac{1}{2 m_1} \right) + 
\frac{(\vsq{p_2}-\vsq{p_1})}{2 m_r} 
\left ( \frac{\vsq{p_2}}{2 m_2} - \frac{\vsq{p_1}}{2 m_1} \right )  \right \} \, .
\end{multline} 
The symmetry condition $S_{k \ell}=S_{\ell k}$, where
\begin{equation}
S_{i j} = \int \dbar{3}{ p_2} \, \dbar{3}{ p_1}  \, 
\psi^\dagger(\vec p_2) \,
(\vsq{p_2})^k \,
(\vsq{p_1})^\ell \,
\frac{\lambda^2}{ \vec k\,^2 \Omega^4} \psi(\vec p_1) \, ,
\end{equation}
implies that the term in Eq.~\eqref{CC1_form3} proportional to $E
(\vsq{p_2}-\vsq{p_1})$ vanishes, and the final term simplifies, so that
\begin{equation} \label{result_CC1}
\Delta E_{CC1} = (-4 \pi Z \alpha) \frac{\alpha}{\pi} 
\int_0^1 \dd v \, f_1(v) \int  \dbar{3}{p_2} \, \dbar{3}{p_1}  \, 
\psi^\dagger(\vec p_2) 
\left \{ 2 s^{-1}(\vec p_2) s^{-1}(\vec p_1) + 
\frac{(\vsq{p_2}-\vsq{p_1})^2}{8 m_r^2} \right \} 
\frac{\lambda^2}{ \vec k\,^2 \Omega^4}  \psi(\vec p_1) \, .
\end{equation}
The $\calO[\alpha (Z \alpha)^4 m_r])$ contribution from single
VP-corrected Coulomb exchange (Fig.~\ref{fig1}(a)),
from (\ref{single_Coulomb_exchange}) and (\ref{expan_of_R_for_C}),
is
\begin{equation} \label{result_C2}
\Delta E_{C2} = (-4 \pi Z \alpha) \frac{\alpha}{\pi} 
\int_0^1 \dd v \, f_1(v) \int  \dbar{3}{p_2} \, \dbar{3}{p_1}  \, 
\psi^\dagger(\vec p_2) 
\left \{ - s^{-1}(\vec p_2) s^{-1}(\vec p_1) - 
\frac{(\vsq{p_2}-\vsq{p_1})^2}{8} 
\left ( \frac{1}{m_2^2} + \frac{1}{m_1^2} \right ) \right \} 
\frac{\lambda^2}{ \vec k\,^2 \Omega^4}  \psi(\vec p_1) \, .
\end{equation}
The total $\calO[\alpha (Z \alpha)^4 m_r]$ contribution, from 
\eqref{result_M}, 
\eqref{result_CCC0}, 
\eqref{result_CC1}, and
\eqref{result_C2}, is
\begin{multline}
\label{cancel2}
\Delta E_{\rm tot} = 
\delta \Delta E_{M}  
+ \Delta E_{CCC0} 
+ \Delta E_{CC1} 
+ \Delta E_{C2} 
= (-4 \pi Z \alpha) \frac{\alpha}{\pi} 
\int_0^1 \dd v \, f_1(v) \int  \dbar{3}{p_2} \, 
\dbar{3}{p_1}  \, \psi^\dagger(\vec p_2) 
\frac{\lambda^2}{ \vec k\,^2 \Omega^4}  \psi(\vec p_1) \\
\times 
\bigg \{ 
- \bigg[ \frac{(\vsq{p_2}-\vsq{p_1})^2}{8} 
\left ( \frac{2}{m_1 m_2} \right ) \bigg ] 
- \bigg [ s_2^{-1} s_1^{-1} \bigg ] 
+ \bigg [ 2 s_2^{-1} s_1^{-1} + 
\frac{(\vsq{p_2}-\vsq{p_1})^2}{8 m_r^2} \bigg ]
-\bigg [ s_2^{-1} s_1^{-1} +
\frac{(\vsq{p_2}-\vsq{p_1})^2}{8} 
\left ( \frac{1}{m_2^2} + \frac{1}{m_1^2} \right ) \bigg ]\bigg \} = 0 \, ,
\end{multline}

\end{widetext}
where $s_i \equiv s(\vec p_i)$.  We see that the total difference between the
energy corrections in the RC gauge and OC gauge at order $\alpha (Z
\alpha)^4 m_r$ is zero, confirming the gauge invariance of the energy levels at
this order.
A remark is in order. Namely, the NRQED$_\mu$ Lagrangian 
given in Eq.~\eqref{Lagrangian_outline} is constructed,
{\em a priori}, for spin-$\half$ orbiting particles,
and spin-$\half$ nuclei. 
However, the cancellation mechanism for the gauge-dependent
terms turns out to be manifestly independent of the nuclear spin.
This is because we have only used the spin-independent Coulomb coupling
and the magnetic photon exchange.
Both of these terms are independent of the muon
spin, and of the nuclear spin.
For the generalization of Eq.~\eqref{Lagrangian_outline}
to heavy particles (whose mass is commensurable or
exceeds the muon mass), one can use 
Eqs.~(1)--(4) of Ref.~\cite{ZaPa2010},
or Eq.~(12.99) of Ref.~\cite{JeAd2022book}.
One realizes that the Coulomb coupling $q_i \, A^0$ 
and the coupling of the convection current to 
the vector potential, manifest in the term
$\vec\pi^2 = (\vec p - q_i \vec A)^2 \sim
-q_i \, (\vec p \cdot \vec A + \vec A \cdot \vec p)$, 
are independent of the particle's spin. 
Because these coupling are the only ones that 
enter the cancellation mechanism that leads to 
gauge invariance, our derivation remains valid 
for arbitrary nuclear spin. 

\vspace{-0.4cm}

\section{Numerical Values for Selected Bound Systems}
\label{sec5}
\vspace{-0.2cm}

After discussing the cancellation 
of the gauge-dependent terms,
we now turn to 
numerical calculations of
theoretical predictions for the 
gauge-invariant terms, 
following 
the methods developed in
Refs.~\cite{Pa1996mu,Je2011pra,KaIvKo2012}.
We have extended the calculations to include additional states and
muonic ions that have not been previously considered.

Some of these ions have nuclei with integer spin. 
A derivation of the relativistic and recoil corrections 
to vacuum polarization in muonic bound systems 
for nuclei with arbitrary spin 
has been given in Ref.~\cite{Je2011pra}.
The derivation is 
based on the generalization of the 
Breit Hamiltonian for interacting 
particles with arbitrary spin 
[see Eq.~(12.100) of Ref.~\cite{JeAd2022book}]
for massive photon exchange (photon mass 
$\lambda$). One then integrates over the 
spectral parameter of vacuum polarization 
according to Eq.~\eqref{disp}.
The same expressions can be obtained
by generalizing the lowest-order terms of the 
NRQED$_\mu$ Lagrangian
to include a nucleus of arbitrary spin,
and evaluating the vacuum-polarization
correction to the interaction 
in the OC gauge. 

Indeed, in order to numerically evaluate the corrections in the OC gauge,
we first observe that the nonrelativistic
Hamiltonian is separable.
The nonrelativistic (Schr\"{o}dinger) Hamiltonian 
in the center-of-mass system, where the muon and the nucleus
have opposite momenta $\vec p$ and $-\vec p$, respectively, reads
(in natural units, $\hbar = c = \epsilon_0 = 1$),
\begin{align}
H =& \; \frac{\vecpt}{2 m_\mu} + \frac{\vecpt}{2 m_N}  - \frac{Z\alpha}{r} =
\frac{\vecpt}{2 m_r} - \frac{Z\alpha}{r} \,,
\\
m_r =& \; \frac{m_\mu}{1 + \xi_N} \,,
\qquad
\xi_N = \frac{m_\mu}{m_N} \,.
\end{align}
Here, $m_\mu$ denotes the muon mass,
$m_N$ is the nuclear mass, while $m_r$ is the reduced mass of the 
one-muon ion. This equation can be solved exactly in terms of Schr\"{o}dinger
eigenstates, which have a nonrelativistic radial 
part and an angular part (see Chap.~6 of Ref.~\cite{JeAd2022book}). 

For later use, it turns out to be advantageous to define
the recoil parameter
\begin{equation}
\beta_N = \frac{m_e}{(Z \alpha \, m_r)} \,,
\end{equation}
where $m_e$ is the electron mass, $Z$ is the nuclear charge, and
$\alpha$ is the fine-structure constant.
For muonic hydrogen (\muH), muonic deuterium (\muD),
muonic helium-3 (\muHeThree), muonic helium-4 (\muHeFour),
muonic carbon-12 (\muCTwelve), and muonic carbon-13 (\muCThirteen),
the results read as follows [see also Eq.~\eqref{beta}],
\begin{subequations}
\begin{align}
\beta_{\rm \mu H} =& \; 0.737\,383\,7\ldots \,,
\\
\beta_{\rm \mu D} =& \; 0.700\,086\,1\ldots \,,
\\
\beta_{\rm \mu {}^3 He} =& \; 0.343\,842\,9\dots \,,
\\
\beta_{\rm \mu {}^4 He} =& \; 0.340\,769\,1\dots \,,
\\
\beta_{\rm \mu {}^{12} C} =& \; 0.111\,502\,9\dots \,,
\\
\beta_{\rm \mu {}^{13} C} =& \; 0.111\,422\,4\dots \,,
\end{align}
\end{subequations}
These numerical values were derived from
the atomic masses compiled in  Ref.~\cite{KoEtAl2021}.
(By a subtraction of the electron rest masses,
the atomic masses of Ref. \cite{KoEtAl2021} yield the nuclear masses to the required 
accuracy.)

Our extensive considerations
on gauge invariance have led to the 
conclusion that the relativistic and 
recoil corrections to vacuum polarization,
in the order $\alpha (Z\alpha)^4 m_r$,
can be evaluated based on the 
nonretarded approximation to the 
optimized Coulomb gauge.
In the optimized Coulomb gauge, one 
starts from the vacuum-polarization
corrected photon propagator given in 
Eq.~(\ref{OC_propagator}), and expands to first order in 
the vacuum-polarization scalar $\Pi(k\,^2)$, 
which itself is expanded up to $\calO(\alpha)$.
The one-loop VP correction
to the time-time component of the 
Coulomb-gauge photon propagator is expressed as a 
parametric integral, leading to
\begin{equation}
D_{00}^{\rm OC}(k) = \frac{1}{\vec k\,^2}
+ \frac{\alpha}{\pi} 
\int_0^1 dv \, f_1(v) \frac{1}{\vec k\,^2+\lambda^2} + \cdots \, ,
\end{equation}
where $\lambda = 2 m_e/\sqrt{1-v^2}$.

From the time-time component 
of the photon propagator, one thus 
obtains the leading nonretarded one-loop VP correction
to energy levels, which can be 
expressed in terms of an 
integration over a Yukawa potential
$v_\vp(\lambda; r)$, which we define as
\begin{equation}
v_\vp(\lambda; r) = -\frac{Z\alpha}{r} \, \ee^{-\lambda \, r} \,.
\end{equation}
We define the linear integration operator $K$,
\begin{equation}
V_\vp(r) = K[ v_\vp(m_e \, \rho; r) ] \,,
\qquad \lambda = m_e \, \rho \,,
\end{equation}
where
\begin{equation}
K[f(\rho)] = \frac{2 \alpha}{3 \pi} \int\limits_2^\infty \dd \rho \;
 \frac{2 + \rho^2 }{\rho^3} \sqrt{1 - \frac{4}{\rho^2}}  \; f(\rho) \,.
\end{equation}
We use non-relativistic Schr\"{o}dinger states
$| \psi_{n \ell_j} \rangle$
(see Sec.~6.4 of Ref.~\cite{JeAd2022book}),
where $n$ is the principal quantum number,
$\ell$ is the orbital angular momentum quantum number,
and $j$ is the total angular momentum quantum number,
noting that the results are independent of the 
magnetic projection quantum number.
One calculates the matrix element 
$\langle \psi_{n \ell_j} | v_\vp(\lambda; r)
| \psi_{n \ell_j} \rangle$
and obtains the leading vacuum-polarization 
energy correction as
\begin{equation}
E_\vp = \langle \psi_{n \ell_j} | V_\vp(r) | \psi_{n \ell_j} \rangle 
= K[ \langle \psi_{n \ell_j} | v_\vp(\lambda; r)
| \psi_{n \ell_j} \rangle ].
\end{equation}

Matching the scattering amplitude with the effective Hamiltonian,
one can then obtain, in the optimized Coulomb gauge,
the relativistic and recoil corrections 
due to vacuum-polarization in terms 

\begin{equation}
\label{EE1}
\delta E^{(1)} = 
K\left[ \left< \psi_{n \ell_j}\left| \delta \calH \right| \psi_{n \ell_j} \right> \right] \,,
\end{equation}
where $\delta  \calH$ is the sum of four terms,
\begin{subequations}
\begin{equation}
\delta \calH = \sum_{j=1}^{4} \delta \calH_j \,.
\end{equation}
Some of the terms depend on the nuclear spin (see Ref.~\cite{PaKa1995}),
and it is advantageous to define
\begin{equation}
\delta_I = \left\{
\begin{array}{cc}  1  & \qquad \mbox{half-integer nuclear spin} \\
0  & \qquad \mbox{integer nuclear spin}
\end{array} \right. \,.
\end{equation}
The terms $\delta \calH_j$ can be broken down as follows 
\begin{align}
\delta \calH_1 = & \; 
\frac{Z\alpha}{8} \left( \frac{1}{m_\mu^2} + \frac{\delta_I}{m_N^2} \right) \,
\left( 4 \pi \delta^3(\vec r\,) - \frac{\lambda^2 }{r} \, \ee^{-\lambda r} \right) \,,
\\[0.11ex]
\delta \calH_2 =& \; -\frac{Z\alpha \lambda^2 \ee^{-\lambda r}}{4 m_\mu m_N r}\,
\left( 1 - \frac{\lambda\,r}{2} \right) \,,
\\[0.11ex]
\delta \calH_3 =& \; -\frac{Z\alpha }{2 m_\mu m_N} \;
p^i \; \ee^{-\lambda r} \; \left( \frac{\delta^{ij}}{r} + \frac{1 + \lambda\, r}{r^3}  \, r^i r^j
\right) \; p^j \,,
\\[0.11ex]
\delta \calH_4 =& \; Z\alpha \left( \frac{1}{4 m_\mu^2} + \frac{1}{2 m_\mu m_N} \right)
\frac{\ee^{-\lambda r} \, (1 + \lambda r)}{r^3} \; \vec \sigma \cdot \vec L \,,
\end{align}
\end{subequations}
where we use the summation convention for the 
superscripts $i$ and $j$ which denote the Cartesian 
components of the position and momentum operators.
This form for $\delta \calH$ agrees with that used in earlier work 
\cite{Je2011pra}.  We take the opportunity to correct the prefactor to
$\delta w_3$ in Eq.~(24c) of Ref.~\cite{Je2011pra}: The expression
$4 m_\mu m_N$ in the denominator of the prefactor should be replaced
by $2 m_\mu m_N$.  Also, the left-most $p^i$ of Eq.~(24c) should be on
the left of that term instead of to the right of the $e^{-\lambda r}$ factor.

However, we should remember that the reference state
$| \psi_{n \ell_j} \rangle$ also receives relativistic 
corrections, and these lead to further 
energy shifts in view of second-order 
matrix elements. The relevant terms in the Breit 
Hamiltonian read as follows,
\begin{align}
\label{deltaH}
\delta H =& \; \sum_{j=1}^{4} \delta H_j \,,
\qquad
\delta H_1 =  - \frac{\vec p^{\,4}}{8 m_\mu^3 } - \frac{\vec p^{\,4}}{8 m_N^3 }  \,,
\nonumber\\
\delta H_2 =& \;  \left( \frac{1}{m_\mu^2} + \frac{\delta_I}{m_N^2} \right) 
\frac{\pi Z \alpha \, \delta^3(\vec r\,)}{2} \,,
\nonumber\\
\delta H_3 =& \; -\frac{Z\alpha}{2 m_\mu m_N} p^i \left( \frac{\delta^{i j}}{r} + 
\frac{r^i \, r^j}{r^3} \right) p^j \,,
\nonumber\\
\delta H_4  =& \;  \frac{Z\alpha}{r^3} \,
\left( \frac{1}{4 m_\mu^2} + \frac{1}{2 m_\mu m_N} \right)
\; \vec \sigma \cdot \vec L \,.
\end{align}
In view of corrections to both bra and ket vectors, one
incurs a factor two,
\begin{equation}
\label{EE2}
\delta E^{(2)} = 2 \, K\left[ \left< \psi_{n \ell_j} \left|
\delta H \right| \delta \psi_{n \ell_j} \right> \right] \,,
\end{equation}
for the relativistic wave-function correction.
(Here, $ \delta \psi_{n j_\ell} $ is the one-loop 
vacuum-polarization-induced wave function
correction [see Ref.~\cite{Je2011pra}].)
The $\alpha (Z\alpha)^4 \, m_r$ relativistic recoil
correction to vacuum polarization is finally found to be 
\begin{equation}
\label{deltaEvp}
\delta E_\vp = \delta E^{(1)} + \delta E^{(2)} \,,
\end{equation}
where we appeal to Eqs.~\eqref{EE1} and~\eqref{EE2}.
Numerical values of the corrections are presented in Table~\ref{table1}.
We include values for muonic carbon,
which has been studied quite intensely in this
context~\cite{ScEtAl1982,RuEtAl1984}
(for scattering data, see Ref.~\cite{OfEtAl1991}).

Some remarks on the comparison with other 
results reported in the literature are in order.
Our values for the Lamb shift of $n=2$
states in \muH, \muD, \muHeThree,
and \muHeFour{} are consistent with prior work, specifically with
those from Refs.~\cite{Je2011pra,KaIvKo2012}.
Values for the $n=2$ fine structure 
interval $2P_{3/2}$--$2P_{1/2}$ (of the
ions listed in Table~\ref{table1}) are consistent with Ref.~\cite{KaIvKo2012}.
For fine-structure and Lamb-shift intervals,
our values are also in reasonable numerical
agreement (better than 2\%) with results given in
Ref.~\cite{Bo2011preprint,Bo2012,ElKrMa2011,FrEtAl2017,DiEtAl2018}.
For the $2S$--$1S$ interval in~\muH, we get 
(0.14285-0.02425) meV = 0.11860 meV, while Dorokhov {\it et al.}
(Ref.~\cite{DoEtAl2019}) find 0.1263 meV, which signals
agreement to better than 6.5\%.

% *** TABLE ***
\begin{table*}[th!]
\caption{\label{table1} The relativistic and recoil
correction to vacuum polarization,
$\delta E_\vp$ defined in Eq.~\eqref{deltaEvp}, is presented
for muonic hydrogen and deuterium,
muonic helium ions, and muonic carbon ions
for reference states with
principal quantum numbers $n \leq 4$.
All energies are given in units of meV.
The results given are correct to within $\pm 1$ in the least 
significant figure shown, and are in agreement with prior work
(as discussed in \cite{Je2011pra}).
}
\renewcommand{\arraystretch}{1.2}
\begin{tabular}{l@{\hspace{0.2cm}}l@{\hspace{0.2cm}}l@{\hspace{0.2cm}}l@{\hspace{0.2cm}}l@{\hspace{0.2cm}}l@{\hspace{0.2cm}}l@{\hspace{0.2cm}}l}
\hline
\hline
 & \multicolumn{7}{c}{\muH{}\;(all entries in meV)} \\
 & \multicolumn{1}{c}{$S_{1/2}$} & \multicolumn{1}{c}{$P_{3/2}$} 
 & \multicolumn{1}{c}{$D_{5/2}$} & \multicolumn{1}{c}{$F_{7/2}$} 
 & \multicolumn{1}{c}{$P_{1/2}$} & \multicolumn{1}{c}{$D_{3/2}$} 
 & \multicolumn{1}{c}{$F_{5/2}$} \\
\hline
$n = 1$ & $ -1.4285 \times 10^{-1} $ & \multicolumn{1}{c}{---} & \multicolumn{1}{c}{---} & \multicolumn{1}{c}{---} & \multicolumn{1}{c}{---} & \multicolumn{1}{c}{---} & \multicolumn{1}{c}{---} \\
$n = 2$ & $ -2.4245 \times 10^{-2} $ & $ -5.2262 \times 10^{-4} $ & \multicolumn{1}{c}{---} & \multicolumn{1}{c}{---} & $ -5.4864 \times 10^{-3} $ & \multicolumn{1}{c}{---} & \multicolumn{1}{c}{---} \\
$n = 3$ & $ -7.1344 \times 10^{-3} $ & $ -2.1965 \times 10^{-4} $ & $ -2.9404 \times 10^{-6} $ & \multicolumn{1}{c}{---} & $ -1.8540 \times 10^{-3} $ & $ -2.0700 \times 10^{-5} $ & \multicolumn{1}{c}{---} \\
$n = 4$ & $ -2.9494 \times 10^{-3} $ & $ -1.0054 \times 10^{-4} $ & $ -2.0114 \times 10^{-6} $ & $ -1.5354 \times 10^{-8} $ & $ -8.1028 \times 10^{-4} $ & $ -1.2171 \times 10^{-5} $ & $ -9.4409 \times 10^{-8} $ \\
\hline
 & \multicolumn{7}{c}{\muD{}\;(all entries in meV)} \\
 & \multicolumn{1}{c}{$S_{1/2}$} & \multicolumn{1}{c}{$P_{3/2}$} 
 & \multicolumn{1}{c}{$D_{5/2}$} & \multicolumn{1}{c}{$F_{7/2}$} 
 & \multicolumn{1}{c}{$P_{1/2}$} & \multicolumn{1}{c}{$D_{3/2}$} 
 & \multicolumn{1}{c}{$F_{5/2}$} \\
\hline
$n = 1$ & $ -1.6541 \times 10^{-1} $ & \multicolumn{1}{c}{---} & \multicolumn{1}{c}{---} & \multicolumn{1}{c}{---} & \multicolumn{1}{c}{---} & \multicolumn{1}{c}{---} & \multicolumn{1}{c}{---} \\
$n = 2$ & $ -2.8149 \times 10^{-2} $ & $ -6.3147 \times 10^{-4} $ & \multicolumn{1}{c}{---} & \multicolumn{1}{c}{---} & $ -6.3675 \times 10^{-3} $ & \multicolumn{1}{c}{---} & \multicolumn{1}{c}{---} \\
$n = 3$ & $ -8.2997 \times 10^{-3} $ & $ -2.6724 \times 10^{-4} $ & $ -3.9686 \times 10^{-6} $ & \multicolumn{1}{c}{---} & $ -2.1473 \times 10^{-3} $ & $ -2.6177 \times 10^{-5} $ & \multicolumn{1}{c}{---} \\
$n = 4$ & $ -3.4359 \times 10^{-3} $ & $ -1.2252 \times 10^{-4} $ & $ -2.7168 \times 10^{-6} $ & $ -2.3162 \times 10^{-8} $ & $ -9.3765 \times 10^{-4} $ & $ -1.5370 \times 10^{-5} $ & $ -1.3098 \times 10^{-7} $ \\
\hline
 & \multicolumn{7}{c}{\muHeThree{}\;(all entries in meV)} \\
 & \multicolumn{1}{c}{$S_{1/2}$} & \multicolumn{1}{c}{$P_{3/2}$} 
 & \multicolumn{1}{c}{$D_{5/2}$} & \multicolumn{1}{c}{$F_{7/2}$} 
 & \multicolumn{1}{c}{$P_{1/2}$} & \multicolumn{1}{c}{$D_{3/2}$} 
 & \multicolumn{1}{c}{$F_{5/2}$} \\
\hline
$n = 1$ & $ -4.6564 \times 10^{0} $ & \multicolumn{1}{c}{---} & \multicolumn{1}{c}{---} & \multicolumn{1}{c}{---} & \multicolumn{1}{c}{---} & \multicolumn{1}{c}{---} & \multicolumn{1}{c}{---} \\
$n = 2$ & $ -8.2167 \times 10^{-1} $ & $ -4.3125 \times 10^{-2} $ & \multicolumn{1}{c}{---} & \multicolumn{1}{c}{---} & $ -3.1232 \times 10^{-1} $ & \multicolumn{1}{c}{---} & \multicolumn{1}{c}{---} \\
$n = 3$ & $ -2.4087 \times 10^{-1} $ & $ -1.7558 \times 10^{-2} $ & $ -8.0660 \times 10^{-4} $ & \multicolumn{1}{c}{---} & $ -9.8676 \times 10^{-2} $ & $ -3.5897 \times 10^{-3} $ & \multicolumn{1}{c}{---} \\
$n = 4$ & $ -9.9322 \times 10^{-2} $ & $ -7.9110 \times 10^{-3} $ & $ -5.1987 \times 10^{-4} $ & $ -1.5665 \times 10^{-5} $ & $ -4.2176 \times 10^{-2} $ & $ -1.9750 \times 10^{-3} $ & $ -5.6537 \times 10^{-5} $ \\
\hline
 & \multicolumn{7}{c}{\muHeFour{}\;(all entries in meV)} \\
 & \multicolumn{1}{c}{$S_{1/2}$} & \multicolumn{1}{c}{$P_{3/2}$} 
 & \multicolumn{1}{c}{$D_{5/2}$} & \multicolumn{1}{c}{$F_{7/2}$} 
 & \multicolumn{1}{c}{$P_{1/2}$} & \multicolumn{1}{c}{$D_{3/2}$} 
 & \multicolumn{1}{c}{$F_{5/2}$} \\
\hline
$n = 1$ & $ -4.7546 \times 10^{0} $ & \multicolumn{1}{c}{---} & \multicolumn{1}{c}{---} & \multicolumn{1}{c}{---} & \multicolumn{1}{c}{---} & \multicolumn{1}{c}{---} & \multicolumn{1}{c}{---} \\
$n = 2$ & $ -8.4040 \times 10^{-1} $ & $ -4.4276 \times 10^{-2} $ & \multicolumn{1}{c}{---} & \multicolumn{1}{c}{---} & $ -3.1930 \times 10^{-1} $ & \multicolumn{1}{c}{---} & \multicolumn{1}{c}{---} \\
$n = 3$ & $ -2.4645 \times 10^{-1} $ & $ -1.8056 \times 10^{-2} $ & $ -8.4033 \times 10^{-4} $ & \multicolumn{1}{c}{---} & $ -1.0083 \times 10^{-1} $ & $ -3.7153 \times 10^{-3} $ & \multicolumn{1}{c}{---} \\
$n = 4$ & $ -1.0165 \times 10^{-1} $ & $ -8.1385 \times 10^{-3} $ & $ -5.4174 \times 10^{-4} $ & $ -1.6579 \times 10^{-5} $ & $ -4.3088 \times 10^{-2} $ & $ -2.0427 \times 10^{-3} $ & $ -5.9346 \times 10^{-5} $ \\
\hline
 & \multicolumn{7}{c}{\muCTwelve{}\;(all entries in meV)} \\
 & \multicolumn{1}{c}{$S_{1/2}$} & \multicolumn{1}{c}{$P_{3/2}$} 
 & \multicolumn{1}{c}{$D_{5/2}$} & \multicolumn{1}{c}{$F_{7/2}$} 
 & \multicolumn{1}{c}{$P_{1/2}$} & \multicolumn{1}{c}{$D_{3/2}$} 
 & \multicolumn{1}{c}{$F_{5/2}$} \\
\hline
$n = 1$ & $ -7.0348 \times 10^{2} $ & \multicolumn{1}{c}{---} & \multicolumn{1}{c}{---} & \multicolumn{1}{c}{---} & \multicolumn{1}{c}{---} & \multicolumn{1}{c}{---} & \multicolumn{1}{c}{---} \\
$n = 2$ & $ -1.4104 \times 10^{2} $ & $ -1.5300 \times 10^{1} $ & \multicolumn{1}{c}{---} & \multicolumn{1}{c}{---} & $ -8.8894 \times 10^{1} $ & \multicolumn{1}{c}{---} & \multicolumn{1}{c}{---} \\
$n = 3$ & $ -4.0682 \times 10^{1} $ & $ -6.1742 \times 10^{0} $ & $ -9.2963 \times 10^{-1} $ & \multicolumn{1}{c}{---} & $ -2.5562 \times 10^{1} $ & $ -3.1888 \times 10^{0} $ & \multicolumn{1}{c}{---} \\
$n = 4$ & $ -1.6603 \times 10^{1} $ & $ -2.7148 \times 10^{0} $ & $ -5.2139 \times 10^{-1} $ & $ -7.3599 \times 10^{-2} $ & $ -1.0526 \times 10^{1} $ & $ -1.4693 \times 10^{0} $ & $ -1.9701 \times 10^{-1} $ \\
\hline
 & \multicolumn{7}{c}{\muCThirteen{}\;(all entries in meV)} \\
 & \multicolumn{1}{c}{$S_{1/2}$} & \multicolumn{1}{c}{$P_{3/2}$} 
 & \multicolumn{1}{c}{$D_{5/2}$} & \multicolumn{1}{c}{$F_{7/2}$} 
 & \multicolumn{1}{c}{$P_{1/2}$} & \multicolumn{1}{c}{$D_{3/2}$} 
 & \multicolumn{1}{c}{$F_{5/2}$} \\
\hline
$n = 1$ & $ -7.0409 \times 10^{2} $ & \multicolumn{1}{c}{---} & \multicolumn{1}{c}{---} & \multicolumn{1}{c}{---} & \multicolumn{1}{c}{---} & \multicolumn{1}{c}{---} & \multicolumn{1}{c}{---} \\
$n = 2$ & $ -1.4122 \times 10^{2} $ & $ -1.5321 \times 10^{1} $ & \multicolumn{1}{c}{---} & \multicolumn{1}{c}{---} & $ -8.9012 \times 10^{1} $ & \multicolumn{1}{c}{---} & \multicolumn{1}{c}{---} \\
$n = 3$ & $ -4.0737 \times 10^{1} $ & $ -6.1843 \times 10^{0} $ & $ -9.3148 \times 10^{-1} $ & \multicolumn{1}{c}{---} & $ -2.5597 \times 10^{1} $ & $ -3.1948 \times 10^{0} $ & \multicolumn{1}{c}{---} \\
$n = 4$ & $ -1.6626 \times 10^{1} $ & $ -2.7194 \times 10^{0} $ & $ -5.2251 \times 10^{-1} $ & $ -7.3802 \times 10^{-2} $ & $ -1.0540 \times 10^{1} $ & $ -1.4721 \times 10^{0} $ & $ -1.9751 \times 10^{-1} $ \\
\hline
\hline
\end{tabular}
\end{table*}

\vspace{-0.3cm}

\section{Conclusions}
\label{sec6}
\vspace{-0.2cm}

In this article, we have clarified the 
mechanism behind the gauge invariance 
of the relativistic and recoil corrections
to the vacuum-polarization mediated energy 
shift in muonic atoms. 
We have contrasted the renormalized 
Coulomb gauge given in Eq.~\eqref{RC_photon_propagator}
with the optimized Coulomb gauge 
given in Eq.~\eqref{OC_gauge}.
The leading vacuum-polarization energy 
correction in muonic atoms is of order
$\alpha (Z\alpha)^2 m_r$. 
In order to show gauge invariance
up to the order $\alpha (Z\alpha)^3 m_r$,
we found it sufficient to consider 
one-photon and two-photon exchange diagrams
[see Eq.~\eqref{cancel1}].
However, in the order $\alpha (Z\alpha)^4 m_r$,
we found that it is necessary to include three-photon
exchange diagrams in order to show gauge invariance
[see Eq.~\eqref{cancel2}].
The mechanism described here of attaining
gauge invariance adds the consideration of 
three-photon exchange to the calculations
presented in Ref.~\cite{KaIvKo2012}.

The need for the inclusion
of both two- and three-photon contributions in crossed graphs is expected
by analogy with known results for transverse photon
exchange~\cite{ElKhMi1994,Pa1997prl,Pa1997,CzMeYe1999pra}.
The leading-order contribution of transverse
photon exchange is order $(Z \alpha)^4 m_r$, coming from an instantaneous
approximation to the transverse photon propagator.  A two-photon
graph analogous to Fig.~\ref{fig1}(b), but with the VP-corrected Coulomb 
photon there replaced by a transverse photon, contributes
at order $(Z \alpha)^5 m_r$ (i.e., at next-to-leading order (NLO)), 
and the three-photon graph analogous
to Fig.~\ref{fig1}(c), again with a transverse photon crossing the other two,
contributes at order $(Z \alpha)^6 m_r$ (at next-to-next-to-leading order (NNLO)).  
Both of these crossed graphs have contributions from both soft 
(order $(Z \alpha) m_r $) and ultrasoft (order $(Z \alpha)^2 m_r$) regions 
of the transverse photon three-momentum.  Both soft and ultrasoft
regions make contributions since the denominator of the transverse
photon propagator, $1/k^2 = 1/(k_0^2-\vec k\,^2)$, leads to contributions
where $\vert \vec k\, \vert \sim \vert k_0 \vert$, which  is typically of
order $ (Z \alpha)^2 m_r$ for ultrasoft contributions, in addition to
soft contributions where $\vert \vec k\,\vert \sim (Z \alpha) m_r$.   On the other hand,
the contributions from the VP-corrected Coulomb photon are all
soft, since its propagator denominator is $1/(k^2-\lambda^2) = 1/(k_0^2-\Omega^2)$, 
and $\Omega = \sqrt{\vec k\,^2 + \lambda^2}$ is soft, never ultrasoft.
The soft three-photon-exchange corrections to transverse photon exchange
at NNLO are analogous to the soft three-photon-exchange corrections
to VP-corrected Coulomb photon exchange in the RC gauge at NNLO.
We conclude that three-photon-exchange corrections to 
VP-corrected Coulomb photon exchange in the RC gauge must contribute
at order $\alpha (Z \alpha)^4 m_r $, as was found by the explicit
calculations shown above.

The gauge dependence of recoil corrections to electron vacuum polarization
was considered extensively by Karshenboim {\em et al.} in
Ref.~\cite{KaIvKo2012}.  These authors showed that early evaluations of the
recoil corrections in Refs. \cite{VePa2004,Bo2011preprint,Bo2012}, which
obtained consistent results but which differed from the result
of Ref.~\cite{Je2011pra}, apparently omitted contributing terms due to the 
gauge used.  In Ref.~\cite{KaIvKo2012}, it was shown that two-photon-exchange
graphs must be included when using the renormalized Coulomb gauge
in order to obtain all contributions of order $\alpha (Z \alpha)^4 m_r$. 
We show here that three-photon-exchange graphs also
contribute at this order and must be included in order to confirm
the gauge independence of energy corrections at this order.

Since the modified Coulomb gauge leads to simpler calculations of
corrections, one might wonder why it should not be used immediately with no
further discussion.
Since the standard Coulomb gauge is the traditional default gauge
for Coulombic bound state calculations, we believe it is informative 
to explore the relationship between the standard and optimized
Coulomb gauges and to highlight the problematic effects of a non-instantaneous
time-time photon propagator that appears in the standard Coulomb gauge
when it is renormalized.

Our considerations of gauge 
invariance make use of a revision of NRQED,
due to the different energy scales involved
with bound muons. The hard cutoff scale of NRQED$_\mu$ 
is equal to the muon mass scale, 
while the electron mass scale is approximately 
commensurate with the ``soft'' energy scale representing
the size $\alpha m_\mu$ of typical momenta
in the muonic bound system.
The relativistic electron is therefore
included in NRQED$_\mu$ in terms of a fully 
relativistic field operator
[see Eq.~\eqref{Lagrangian_outline}].
Based on our treatment, employing the OC gauge,
we present numerical results for the 
gauge-independent values of the relativistic and recoil
corrections to VP for one-muon ions
with nuclear charge numbers $Z=1$, $\!2 \!$, and $6$
(see Table~\ref{table1}).

\vspace{-0.4cm}

\section*{Acknowledgments}

\vspace{-0.2cm}

This work was supported by the National Science Foundation through Grants 
PHY-2308792 (G.S.A.) and PHY--2110294 (U.D.J.), and by the National Institute
of Standards and Technology Grant 60NANB23D230 (G.S.A.).


\begin{thebibliography}{10}

\bibitem{Ue1935}
E.~A. Uehling, {\em \relax{Polarization Effects in the Positron Theory}},
  Phys. Rev. {\bf 48},  55  (1935).

\bibitem{JeAd2022book}
U.~D. Jentschura and G.~S. Adkins, {\em \relax{Quantum Electrodynamics: Atoms,
  Lasers and Gravity}} (World Scientific, Singapore, 2022).

\bibitem{LaJe2024}
S. Laporta and U.~D. Jentschura, \relax{Dimensional Regularization and
  Two--loop vacuum polarization operator: Master integrals, analytic results
  and energy shifts}, Phys. Rev. D {\bf 109}, 096020 (2024).

\bibitem{KaIvKa2013}
E.~Y. Korzinin, V.~G. Ivanov, and S.~G. Karshenboim, {\em \relax{$\alpha^2
  (Z\alpha)^{4}m$ contributions to the Lamb shift and the fine structure in
  light muonic atoms}},  Phys. Rev. D {\bf 88},  125019  (2013).

\bibitem{KrEtAl2016}
J. J. Krauth, M. Diepold, B. Franke, A. Antognini, F. Kottmann, and R. Pohl,
  {\em \relax{Theory of the n=2 levels in muonic deuterium}}, Ann. Phys. (N.Y.)
  {\bf 366}, 168-196 (2016).

\bibitem{KaKoShIv2017}
S.~G. Karshenboim, E.~Y. Korzinin, V.~A. Shelyuto, and V.~G. Ivanov, {\em
  \relax{Theory of the Lamb shift in muonic tritium and the muonic $^3\!$He
  ion}},  Phys. Rev. A {\bf 96},  022505  (2017).

\bibitem{FrEtAl2017}
B. Franke, J.~J. Krauth, A. Antognini, M. Diepold, F. Kottmann, and R. Pohl,
  {\em \relax{Theory of the $n=2$ levels in muonic helium-3 ions}},  Eur. Phys.
  J. D {\bf 71},  341  (2017).

\bibitem{ReDi2018}
W.~W. Repko and D.~A. Dicus, {\em \relax{Muonic hydrogen and the proton size}},
   Phys. Rev. D {\bf 98},  013002  (2018).

\bibitem{DiEtAl2018}
M. Diepold, B. Franke, J.~J. Krauth, A. Antognini, F. Kottmann, and R. Pohl,
  {\em \relax{Theory of the Lamb shift and fine structure in muonic $^4\!$He
  ions and the muonic $^3\!$He-$^4\!$He isotope shift}},  Ann. Phys. (N.Y.)
  {\bf 396},  220--244  (2018).

\bibitem{Re2018}
W.~W. Repko and D.~A. Dicus, {\em \relax{Relativistic corrections to vacuum
  polarization contributions in muonic hydrogen}},  Phys. Rev. D {\bf 98},
  033006  (2018).

\bibitem{DoEtAl2019}
A.~E. Dorokhov, A.~P. Martynenko, F.~A. Martynenko, O.~S. Sukhorukova, and
  R.~N. Faustov, {\em \relax{Energy Interval 1S--2S in muonic hydrogen and
  helium}},  JETP {\bf 129},  956--972  (2019).

\bibitem{DoEtAl2020}
A.~E. Dorokhov, R.~N. Faustov, A.~P. Martynenko, and F.~A. Martynenko, {\em
  \relax{Energy interval $3S$--$1S$ in muonic hydrogen}},  Phys. Rev. A {\bf
  102},  062820  (2020).

\bibitem{AnHaPa2022}
A. Antognini, F. Hagelstein, and V. Pascalutsa, {\em \relax{The proton
  structure in and out of muonic hydrogen}},  Annu. Rev. Nucl. Part. Sci. {\bf
  72},  389--418  (2022).

\bibitem{PaEtAl2024}
K. Pachucki, V. Lensky, F. Hagelstein, S.~S. Li~Muli, S. Bacca, and R. Pohl,
  {\em Comprehensive theory of the Lamb shift in light muonic atoms},  Rev.
  Mod. Phys. {\bf 96},  015001  (2024).

\bibitem{BeLiPi1982vol4}
V.~B. Berestetskii, E.~M. Lifshitz, and L.~P. Pitaevskii, {\em \relax{Quantum
  Electrodynamics, Volume 4 of the Course on Theoretical Physics}}, 2 ed.
  (Pergamon Press, Oxford, UK, 1982).

\bibitem{Lo1978}
S. Love, {\em \relax{A Study of Gauge Properties of the Bethe--Salpeter
  Equation for two--fermion electromagnetic bound state systems}},  Ann. Phys.
  (N.Y.) {\bf 113},  153--176  (1978).

\bibitem{BaRe1978}
R. Barbieri and E. Remiddi, {\em \relax{Solving the Bethe-Salpeter equation for
  positronium}},  Nucl. Phys. B {\bf 141},  413--422  (1978).

\bibitem{FeFuHe1980i}
G. Feldman, T. Fulton, and D.~L. Heckathorn, {\em \relax{Gauge invariance of
  relativistic two-particle bound-state energies. 1.}},  Nucl. Phys. B {\bf
  167},  364--382  (1980).

\bibitem{FeFuHe1980ii}
G. Feldman, T. Fulton, and D.~L. Heckathorn, {\em \relax{Gauge invariance of
  relativistic two-particle bound-state energies. 2. Invariant Subsets in
  Perturbation Theory}},  Nucl. Phys. B {\bf 174},  89--108  (1980).

\bibitem{Li1990}
I. Lindgren, {\em \relax{Gauge dependence of interelectronic potentials}},  J.
  Phys. B {\bf 23},  1085--1093  (1990).

\bibitem{Pa1996mu}
K. Pachucki, {\em \relax{Theory of the Lamb shift in muonic hydrogen}},  Phys.
  Rev. A {\bf 53},  2092--2100  (1996).

\bibitem{Je2011pra}
U.~D. Jentschura, {\em \relax{Relativistic reduced--mass and recoil corrections
  to vacuum polarization in muonic hydrogen, muonic deuterium and muonic helium
  ions}},  Phys. Rev. A {\bf 84},  012505  (2011).

\bibitem{KaIvKo2012}
S.~G. Karshenboim, V.~G. Ivanov, and E.~Y. Korzinin, {\em \relax{Relativistic
  recoil corrections to the electron-vacuum-polarization contribution in light
  muonic atoms}},  Phys. Rev. A {\bf 85},  032509  (2012).

\bibitem{VePa2004}
A. Veitia and K. Pachucki, {\em \relax{Nuclear recoil effects in antiprotonic
  and muonic atoms}},  Phys. Rev. A {\bf 69},  042501  (2004).

\bibitem{Bo2011preprint}
E. Borie, e-print arXiv:1103.1772v5 (physics.atom-ph).  The result for the
relativistic and recoil contributions coming from electronic vacuum polarization 
(in Table 1) was corrected in arXiv:1103.1772v7.

\bibitem{Bo2012}
E. Borie, {\em \relax{Lamb shift in light muonic atoms--Revisited}},  Ann.
  Phys. (N.Y.) {\bf 327},  733--763  (2012).

\bibitem{JeSoIvKa1997}
U.~D. Jentschura, G. Soff, V.~G. Ivanov, and S.~G. Karshenboim, {\em
  \relax{Bound $\mu^{+}\mu^{-}$ system}},  Phys. Rev. A {\bf 56},  4483--4495
  (1997).

\bibitem{KaIvJeSo1998}
S.~G. Karshenboim, V.~G. Ivanov, U.~D. Jentschura, and G. Soff, {\em
  \relax{Bound states of the muon-antimuon system: Lifetimes and hyperfine
  splitting}},  J. Exp. Theor. Phys. {\bf 86},  226--236  (1998), [ZhETF {\bf 113}, 409
  (1998)].

\bibitem{KaJeIvSo1998}
S.~G. Karshenboim, U.~D. Jentschura, V.~G. Ivanov, and G. Soff, {\em
  \relax{Next-to-leading and higher order corrections to the decay rate of
  dimuonium}},  Phys. Lett. B {\bf 424},  397--404  (1998).

\bibitem{LaJi2018}
H. Lamm and Y. Ji, {\em \relax{Predicting and discovering true muonium ($\mu^+
  \mu^-$)}},  Eur. Phys. J. Web of Conferences {\bf 181},  01016  (2018).

\bibitem{Je2011aop1}
U.~D. Jentschura, {\em \relax{Lamb shift in muonic hydrogen. ---I. Verification
  and update of theoretical predictions}},  Ann. Phys. (N.Y.) {\bf 326},
  500--515  (2011).

\bibitem{Je2011aop2}
U.~D. Jentschura, {\em \relax{Lamb shift in muonic hydrogen. ---II. Analysis of
  the discrepancy of theory and experiment}},  Ann. Phys. (N.Y.) {\bf 326},
  516--533  (2011).

\bibitem{CaLe1986}
W.~E. Caswell and G.~P. Lepage, {\em \relax{Effective Lagrangians for bound
  state problems in QED, QCD, and other field theories}},  Phys. Lett. B {\bf
  167},  437--442  (1986).

\bibitem{HiLePaSo2013}
R.~J. Hill, G. Lee, G. Paz, and M.~P. Solon, {\em \relax{NRQED Lagrangian at
  order $1/M^4$}},  Phys. Rev. D {\bf 87},  053017  (2013).

\bibitem{We1995i}
S. Weinberg, {\em \relax{The Quantum Theory of Fields, Volume 1: Foundations}}
  (Cambridge University Press, Cambridge, UK, 1995).

\bibitem{Sc1970vol3}
J. Schwinger, {\em \relax{Particles, Sources and Fields (Volume III)}}
  (Addison-Wesley, Reading, MA, 1970).

\bibitem{Cu1960}
R.~E. Cutkosky, {\em \relax{Singularities and discontinuities of Feynman
  amplitudes}},  J. Math. Phys. {\bf 1},  429--433  (1960).

\bibitem{PeSc1995}
M.~E. Peskin and D.~V. Schroeder, {\em \relax{An Introduction to Quantum Field
  Theory}} (Perseus, Cambridge, Massachusetts, 1995).

\bibitem{KaSa1955}
G. K\"{a}ll\'{e}n and A. Sabry, {\em \relax{Fourth order vacuum polarization}},
  Kong. Dan. Vid. Sel. Mat. Fys. Med. {\bf 29},  1--20  (1955).

\bibitem{BaRe1973}
R. Barbieri and E. Remiddi, {\em \relax{Infra-red divergences and adiabatic
  switching. Fourth-order vacuum polarization}},  Nuovo Cim. A {\bf 13},
  99--119  (1973).

\bibitem{EiSh2015}
M.~I. Eides and V.~A. Shelyuto, {\em \relax{Hard three-loop corrections to
  hyperfine splitting in positronium and muonium}},  Phys. Rev. D {\bf 92},
  013010  (2015).

\bibitem{Ad2018}
G.~S. Adkins, {\em \relax{Higher order corrections to positronium energy
  levels}},  J. Phys. Conf. Ser. {\bf 1138},  012005  (2018).

\bibitem{ZaPa2010}
J. Zatorski and K. Pachucki, {\em \relax{Electrodynamics of finite-size
  particles with arbitrary spin}},  Phys. Rev. A {\bf 82},  052520  (2010).

\bibitem{KoEtAl2021}
F.~G. Kondev, M. Wang, W.~J. Huang, S. Naimi, and G. Audi, {\em \relax{The
  NUBASE2020 evaluation of nuclear physics properties}},  Chin. Phys. C {\bf
  45},  030001  (2021).

\bibitem{PaKa1995}
K. Pachucki and S.~G. Karshenboim, {\em \relax{Nuclear-spin-dependent recoil
  correction to the \relax{Lamb} shift}},  J. Phys. B {\bf 28},  L221--L224
  (1995).

\bibitem{ScEtAl1982}
L.~A. Schaller, L. Schellenberg, T.~Q. Phan, G. Piller, A. Ruetschi, and H.
  Schneuwly, {\em \relax{Nuclear Charge radii of the carbon isotopes
  ${}^{12}{\rm C}$, ${}^{13}{\rm C}$ and ${}^{14}{\rm C}$}},  Nucl. Phys. A
  {\bf 379},  523  (1982).

\bibitem{RuEtAl1984}
W. Ruckstuhl, B. Aas, W. Beer, I. Beltrami, K. Bos, P.~F.~A. Goudsmit, H.~J.
  Leisi, G. Strassner, A. Vacchi, F.~W.~N. de~Boer, U. Kiebele, and R. Weber,
  {\em \relax{Precision measurement of the $2p$-$1s$ transition in muonic
  ${}^{12}{\rm C}$: Search for new muon-nucleon interactions or accurate
  determination of the rms nuclear charge radius}},  Nucl. Phys. A {\bf 430},
  685--712  (1984).

\bibitem{OfEtAl1991}
E.~A. J.~M. Offermann, L.~S. Cardman, C.~W. de~Jager, H. Miska, C. de~Vries,
  and H. de~Vries, {\em Energy dependence of the form factor for elastic
  electron scattering from ${}^{12}{\rm C}$},  Phys. Rev. C {\bf 44},
  1096--1117  (1991).

\bibitem{ElKrMa2011}
E.~N. Elekina, A.~A. Krutov, and A.~P. Martynenko, {\em \relax{Fine structure
  of the muonic ${}^4$He ion}},  Phys. Part. Nucl. Lett. {\bf 8},  331--336
  (2011).

\bibitem{ElKhMi1994}
A.~S. Elkhovsky, I.~B. Khriplovich, and A.~I. Mil'stein, {\em
  \relax{Corrections of order $\alpha^4 R_\infty$ to the positronium P
  levels}},  J. Exp. Theor. Phys. {\bf 78},  159--164  (1994).

\bibitem{Pa1997prl}
K. Pachucki, {\em \relax{Recoil effects in positronium energy levels to order
  $\alpha^6$}},  Phys. Rev. Lett. {\bf 79},  4120--4123  (1997).

\bibitem{Pa1997}
K. Pachucki, {\em \relax{Effective Hamiltonian approach to the bound state:
  Positronium hyperfine structure}},  Phys. Rev. A {\bf 56},  297--304  (1997).

\bibitem{CzMeYe1999pra}
A. Czarnecki, K. Melnikov, and A.~S. Yelkhovsky, {\em \relax{Positronium
  $S$-state spectrum: Analytic results at $O(m \, \alpha^6)$}},  Phys. Rev. A
  {\bf 59},  4316--4330  (1999).

\end{thebibliography}
\end{document}